\documentclass[12pt]{article}
\usepackage{epsfig}

\usepackage{amsmath}
\usepackage{amssymb}
\usepackage{euscript}

\begin{document}

\begin{titlepage}
 
\begin{flushright}
\bf TPJU-3/2005
\end{flushright}
\vspace{2mm}

\begin{center}
{\LARGE\bf The Femto-experiment for the LHC:} \\
\vspace{2mm}
{\Large\bf The $W$-boson beams and their targets$^{\star}$} \\
\end{center}

\vspace{5mm}


\begin{center}
{\large\bf  M.~W. Krasny$^{a}$, 
            S.\ Jadach$^{b}$ 
            {\rm and} W.\ P\l{}aczek$^{c}$   
}

\vspace{4mm}
{\em $^a$LPNHE, Pierre and Marie Curie University, Tour 33, RdC, \\
                4, pl. Jussieu, 75005 Paris, France,}\\ 
 \vspace{2mm}
{\em $^b$Institute of Nuclear Physics Polish Academy of Sciences,\\
  ul.\ Radzikowskiego 152, 31-342 Cracow, Poland,}\\ \vspace{2mm}
{\em $^c$Marian Smoluchowski Institute of Physics, Jagiellonian University,\\
   ul.\ Reymonta 4, 30-059 Cracow, Poland.}\\ \vspace{2mm}
  
%
\end{center}

\vspace{5mm}
\begin{abstract}

The LHC has been designed as a collider of proton and ion beams.
However, in its experimental program, which is focused mainly on studies
of high energy transfer collisions of Standard Model point-like
particles, protons and ions will play a backstage role.  For  the majority
of the LHC experimentalists, their role will be confined to providing
standardized, acceleration-process-stable envelopes for tunable-density
and tunable-isospin bunches of Standard Model constituents: quarks and gluons. 
The inter-bunch environment of collisions of these Standard Model particles
specific to hadronic colliders and  absent in the leptonic ones,  
has always been considered as a burden -- an  annoying but unavoidable 
price to pay for increasing    
the collision-energy of point-like particles in the storage rings. 
In this paper we shall argue that such a burden
can be converted into an important merit of the high-energy hadronic colliders 
-- a corner-stone for  a fermi-length-scale ``collision-experiment''
employing the bunches of spectator quarks and gluons as tunable ``femtoscopic''
targets for the  beams of short-living electroweak bosons.
\end{abstract}


\vspace{3mm}
\begin{center}
{\it To be submitted to Physical Review D}
\end{center}
 
\vspace{4mm}
\begin{flushleft}
{\bf TPJU-3/2005\\
     March~2005}
\end{flushleft}  

\vspace{2mm}
\footnoterule
\noindent
{\footnotesize
$^{\star}$The work is partly supported by the program of cooperation 
between the IN2P3 and Polish Laboratories No.\ 05-116, and 
by the EC FP5 Centre of Excellence ``COPIRA'' under the contract
  No.\ IST-2001-37259.
}

\end{titlepage}

\section{Introduction}

Perhaps the most important remaining open-question for 
the quantum-field-theory-based description 
of interactions of the basic building blocks of matter is 
the mechanism which drives the high-energy interactions
of longitudinally and transversely polarized, massive electroweak bosons.
The Higgs model  and its known extensions provide a set of possible  
scenarios for such a mechanism.    
These scenarios are very attractive because they provide an abundant supply  of  
road-maps and the corresponding  navigation rules  which could be 
simply picked-up and  followed while exploring, at the LHC 
collider, the {\it ``Terra incognita''} of 
the TeV-energy-scale interactions of electroweak bosons.
By applying these rules the exploration process is reduced to  
the painstaking  scenario-scrutinizing process. Such a strategy   
has two important merits: it is theoretically well-controlled 
and experimentally well-defined. But is has also a price to pay: 
it involves a  considerable risk of overlooking new, unexpected phenomena.

To minimize such a risk, we would like to advocate a complementary
strategy focusing on  the development of handy experimental tools
for the exploration 
process in parallel to  mastering the discovery-guide recipes.
Obviously, the most desirable tool for  exploring  the high-energy interactions
of the electroweak bosons would be a luminous, polarized, high-energy 
beam of electroweak bosons.
If proton beams of energies exceeding $10^{17}$ GeV were available, secondary
beams of electroweak bosons could be easily formed and used in dedicated
experiments -- in close analogy to fixed target muon-beam experiments
which routinely use beams of unstable particles.
  
The central point of the present and of the forthcoming paper
\cite{forthcoming} is that, 
even if experiments using beams of electroweak bosons
cannot be realized at the macroscopic length-scales, 
the LHC collider offers a reduced-scope, yet unique opportunity  to realize    
them  at the femtoscopic length-scale. 

The luminous, high-energy  LHC collider will be a very efficient factory  
to copiously produce the electroweak bosons. 
Its unique merit, with respect to other machines, is 
that these short-living particles could be observed 
at the LHC collider 
over sufficiently long time for experimental  
``femtoscopic''  studies  of their properties and interactions. 
The electroweak bosons will travel,  for the observers co-moving  with 
the LHC bunches, over the atomic distances of up to $10^{4}$  fm  before decaying. 
This defines the maximal lengths, thus the type,  of possible  targets which, 
if arranged to  co-move with the LHC bunches,  could be 
employed in experimental studies of properties and collisions of the electroweak bosons.
Nature provides only one type of target satisfying the above criteria -- the 
atomic nucleus.

Nuclear beams will play a double role in the proposed scheme.
First of all, they will provide standardized bunches of quarks
and gluons of the 
adjustable isospin,  allowing to tune  the fluxes of the electroweak 
bosons. 
In addition, they will supply the co-moving, hadronic matter
of adjustable length,  which will   
serve in forming the effective targets for the beam of electroweak bosons.

The terms  {\it beam of electroweak bosons} and {\it effective
target} are precisely defined in Sections 3 and 4 of this paper.
They will be used here in direct  analogy to the beam and targets 
being used in the muon scattering experiments. 
The creation of the beam of electroweak bosons in collisions 
of fixed-isospin  partonic-bunches 
is  considered here to be equivalent to the processes of creation of the muon beam 
from the fixed-flux and flavour-composition beams of secondary hadrons.
Hard partonic collisions creating the electroweak boson beam 
are  uncorrelated  with the subsequent collisions of the 
produced $W$-boson beam particles with bunches of  
spectator quarks and gluons because of the 
large Lorentz-$\gamma$ factor   of the colliding partonic bunches, high mass 
of electroweak bosons, and their colourless nature. This factorization is analogous
to the factorization of the production and collisions of  the muon beam.
The length of the target for the electroweak boson beam,  
contrary to the fixed-length target for the muon beam, will vary on 
event-by-event basis depending upon the localization of the space-time volume, 
within the bunch of quarks and gluons, where the beam particle was created.
The average length of the target, however, will be fixed for the fixed atomic
number of the primary beam particles.  This quantity defines the {\it effective-target} length.

The most important difference between using the muon beam in a concrete experiment  
and  using the electroweak boson beam is that the former one  can 
be isolated by the suitable absorber and the beam transport system while
the latter one cannot be fully isolated from the beams of spectator quarks
and gluons,  and from the produced hadrons. Therefore,  special measurement
procedures and unfolding methods must  be invented to filter out, 
as much as possible, the electroweak boson collision signals from 
the noise of ordinary collisions of quarks and gluons.

The discussion of such procedures and methods will be presented in 
 a dedicated paper and is not discussed here. This paper is devoted to  
the beam aspects of the femtoscopic experiment.
We shall pick-up the the beam of charged electroweak bosons, 
namely $W^+$ and $W^-$, and present its properties,  methods of
controlling its intensity, momentum-band, polarization
and the luminosity of the $W$--nucleon collisions.

\section{Gedanken experiment}

If Planck-energy-scale  antineutrino beams were available,  
then configuring a fixed target experiment to 
study the properties and the collisions of polarized  $W$-bosons
could follow directly the examples  of the CERN fixed target experiments which use the  
polarized muon beams. 
Polarized $W$-bosons  produced in collisions of the antineutrinos 
with a  polarized  low-energy electron beam  would travel the distance larger 
than $1$ kilometer before decaying.
Thus, there would be a sufficient  space for: (1) filtering out the annihilation 
processes, (2) selecting a narrow momentum  band  of the produced $W$-beam,
(3) measuring the $W$-boson macroscopic electric current,  
(4) placing a suitable collision target, (5) identifying and measuring 
the products of the $W$-boson  collisions. 
For example,  the inclusive cross section, $\sigma^{\lambda_{in}}_{incl}$, for     
scattering  of polarized $W$-bosons on  nucleons could  be derived from the 
measurements by using the canonical  formula:
\begin{eqnarray}     
N_{incl}^{\lambda_{in}}(s_{Wn}, p_W^{out}) = 
{\cal F}_W^{\lambda_{in}}(s_{Wn}) \; 
\sigma^{\lambda_{in}}_{incl}(s_{Wn}, p_W^{out}) \; 
\rho_t \; 
l_t \;,  
\label{gedanken}
\end{eqnarray}
where $N_{incl}^{\lambda_{in}}(s_{Wn},  p_W^{out})$ is the  observed rate of events 
at the $W$--nucleon centre-of-mass-system (CMS) energy squared $s_{Wn}$,
in which the  momentum of the outgoing $W$-boson is $p_W^{out}$.  
${\cal F}_W^{\lambda_{in}}(s_{Wn})$  is the flux of $W$-bosons
having the polarization  $\lambda_{in}$,  
$\rho_t$ is the target density,  and $l_t$ is the target length.

Applying  this formula to the Planck-energy experimental configuration requires  
two conditions, always taken for granted in the fixed target muon experiments,  
to be fulfilled:
\begin{itemize}
\item
The $W$-boson beam  must be formed before arriving at the position 
of the target.  
\item 
The total distance between the beam creation zone and the exit point of the 
target must be significantly smaller than $ c\tau = \gamma_W c \tau_W$, where  $\gamma_W$ 
is the $W$-boson beam Lorentz factor, $\tau_W$ is the lifetime of the $W$-boson
in its rest frame and $c$ is the speed of light in the vacuum.
\end{itemize}

In our `gedanken experiment' they are fulfilled owing to the selection of  
the point-like annihilation process as the source of the $W$-boson beam and owing 
to the Planck-scale  $\gamma_W$ factor which, in the rest frame of the 
target,  assures macroscopic distances over which the $W$-boson could 
travel before its disintegration. 

Such an experiment cannot be presently realized. 
However, as we shall show in the following section, a reduced scope  
of its functions can be realized at the  femtoscopic scale using the LHC beams.   
The LHC beams,  that  will be employed in configuring such an experiment     
are the proton  and the deuteron beams colliding with  the heavy ion beams.
The collisions of these beams  will be referred to, 
in the following,  as the nucleon--nucleus collisions.

\section{The femto-experiment }
\label{sec:femtoexp}

\subsection{Femto-picture} 
\label{subsec:femtopic}

Given the ratio of the  LHC beam energy 
to the Planck-scale energy,  the LHC experiment to study the $W$-boson 
properties and its collisions must be configured at the fermi-length scale.
It is thus bound to use the nuclear medium to form the beam of $W$-bosons and to 
observe their interactions. The nucleus will  be considered, in  the processes
discussed in this paper,  as a classical object described fully by its static 
parameters.
Its dynamic quantum degrees of freedom can be safely neglected in the following 
discussion.

\begin{figure}
\begin{center}
\leavevmode
\epsfig{file=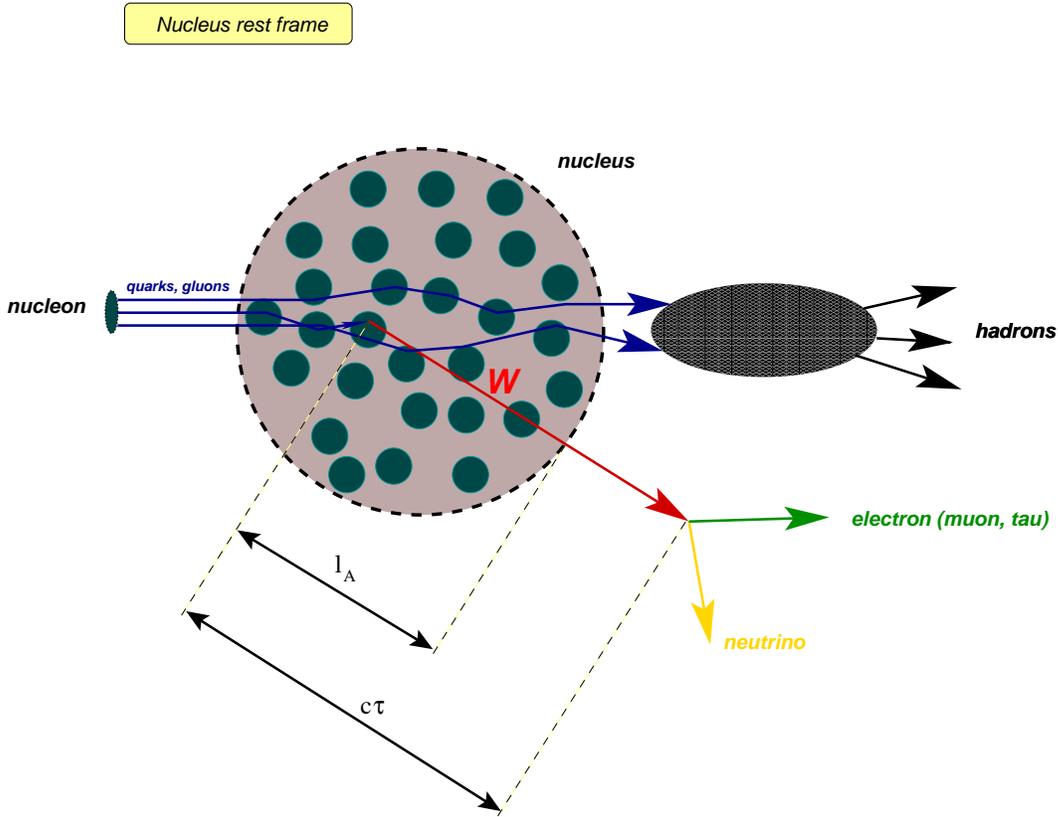,width=14cm}
\end{center}
\caption{\sf The factorization of the $W$-production and $W$-collision processes.}
\label{fig:factorization}
\end{figure}

The nucleus-rest-frame
space-time picture  of production of a $W$-boson in nucleon-nucleus collisions,
its passage in the hadronic matter and subsequent decay 
is  shown in Fig.~\ref{fig:factorization}.
In the nucleus rest frame,  the incoming nucleon 
can be considered, for  the Lorentz factor $\gamma \gg 1$, as 
a Lorentz-frozen (static)  bunch
of independent partons (quarks and gluons). 
The $W$-boson  is produced  in the hard collision of one its  partons with 
the target nucleus. 
The produced $W$-boson travels the distance $l_A$ in the nucleus, then leaves the 
nucleus,  and decays at the distance $ c \tau = c \tau_W \gamma _W$.

Three requirements  must be satisfied to ``configure'' the nucleon--nucleus 
collisions at the LHC as a femtoscopic experiment to explore  the high-energy 
interactions of the $W$-bosons with hadronic matter.
The necessity to fulfill these  requirements determines which of the functions
of the `gedanken experiment'  can be realized at the LHC.
They are discussed below in turn. 

\subsection{Event selection}
\label{subsec:evsel}

The first requirement is to define  the experimental signatures, 
and to provide the effective selection methods 
for  those of the nucleon--nucleus collisions
which contain the signals of the $W$-collisions with hadronic matter.
Contrary to the `gedanken experiment'  
the $W$-boson collision  events cannot be directly selected.
As a consequence the scope of exploring the $W$-collision processes 
must be limited to the inclusive $W$--nucleon collisions,
$W + \textrm{nucleon} \rightarrow W +X$.    
The $W$-boson collision effects,
and the effects of the absorption of the $W$-bosons 
in hadronic matter will be studied  
using the sample of events characterized by  the final-state
$W$-boson signatures. These effects   
will have to be filtered out from the effects 
due to the initial state interaction of the partons producing 
the $W$-bosons, and from the overwhelming ``noise'' 
of simultaneous collisions of 
co-moving  beams of strongly interacting 
partons.  The filtering methods will be proposed and 
their sensitivity to the $W$-collision effects will be discussed 
in \cite{forthcoming}. 

\subsection{Framework} 
\label{secLframework}

\subsubsection{Factorization}
\label{subsec:factorization}

The second requirement is to provide a 
framework, similar to that for the `gedanken experiment', 
which will allow to express  the  nucleon--nucleus collision observables 
exclusively in terms of the $W$--nucleon inclusive cross section and experimentally 
controllable quantities. For simplicity we consider first   the framework for 
the inclusive scattering of unpolarized $W$-bosons,
$W + \textrm{nucleon} \rightarrow W +X$. This framework will be extended 
in  Section~\ref{subsec:frampolar} by including the polarization effects, and in 
Section~\ref{subsec:unfolding} by including  the $W$-boson absorption effects.

The framework is  based upon   a decomposition of the 
process shown in Fig.~\ref{fig:factorization} 
into the three {\bf independent}  processes of:
the creation of the $W$-boson, its propagation in hadronic matter, 
and its decay. We propose the following 
factorized form of  the Born-level matrix element for
the process shown in Fig.~\ref{fig:factorization}:
\begin{equation}
\begin{aligned}
{\cal M} (p_n, p_A, p_W^{in}, r, p_W^{out}, p_1, p_W^{out}-p_1| A) = 
\hspace{30mm}
&\\
{\cal M} _f ( p_n,p_A,p_W^{in} , r  | A)  \; 
{\cal M}_p \left( p_W^{in} , p_W^{out},  l_A( p_W^{in}, p_A , r)\right) \;
&
{\cal M} _d( p_1, p_W^{out}-p_1   ),  
\end{aligned}
\label{eq:master}
\end{equation}
where ${\cal M}_f( p_n,p_A,p_W^{in} , r | A )$
is the amplitude for the $W$-boson formation 
in the collision of the nucleon, carrying the momentum   $p_n$,
with the nucleus of the atomic number $A$,  carrying the momentum $p_A$.
The $W$-boson is created at the position $r$ and carries the  momentum  $p_W^{in}$.  
The  amplitude 
${\cal M}_p\left(p_W^{in} , p_W^{out},  l_A( p_W^{in} ,p_A, r) \right)$
represents  the propagation amplitude  of the produced $W$-boson in hadronic matter 
over the distance $l_A$
leading to a change of its momentum to  $p_W^{out} $. 
This amplitude includes undisturbed propagation of the $W$-boson -
in such a case  $p_W^{in} = p_W^{out}$. 
The amplitude  ${\cal M}_d( p_1, p_W^{out}-p_1   )$
is the decay amplitude  of observing the $SU(2)_L$-doublet particles 
of the momenta $p_1$ and $ p_W^{out}-p_1 $ in the decay of 
the $W$-boson.
In the above formulas the symbols representing kinematic and space 
variables are {\bf three-vectors}. 

Four  properties of the above factorization scheme are important.
\begin{itemize}
\item  
The $W$-boson propagation amplitude,  ${\cal M}_p$,  
does not depend upon 
the mechanism of the $W$-boson formation. It depends only  upon 
its  momentum.
\item
${\cal M}_p$ 
depends upon the position of the $W$-boson creation point, $r$,   only 
via the $W$-boson momentum dependent
effective path-length of the $W$-boson in the hadronic matter,  $l_A$.
\item
The decay amplitude,  ${\cal M}_d$,  is independent 
of the $W$-boson creation point   and of  the atomic number of the nucleus.
It depends only upon the momentum  of the $W$-boson exiting the hadronic matter
which, in turn,  is unambiguously determined by the momenta of
its decay products.
\item
The momenta of the incoming and of the outgoing $W$-boson are 
described by three-vectors rather than by four-vectors. 
\end{itemize}
Such a simple factorization does not work in general. It requires 
several conditions to be satisfied. 
The reasons why it works 
for $W$-bosons produced at the LHC collider are  discussed below.

\subsubsection{The case of $W$-bosons at LHC}

The  factorization is a direct consequence of the following  facts:
\begin{itemize}
\item
The $W$-boson is a point-like particle.  
\item
The $W$-boson is a colour-neutral particle%
  \footnote{While the former condition assures that the formation of the $W$-boson  
            is an instantaneous process, the present one assures that a 
            colour-neutral object is instantaneously produced. Note,  that none of the 
            above two conditions is fulfilled in production of the $J/\Psi$ particle.
            }.        
\item
The mass of the $W$-boson is substantially larger  than any  
scale of the strong interactions.   
\item
The LHC energy is sufficiently high for the $W$-boson decay length to be larger than 
the total distance  between the $W$-boson creation point  and the point  where it exits 
the nuclear medium. 
\item 
The $W$-bosons arriving at the position of the target can be 
considered as free on-shell particles. 
\end{itemize}
The first  three  of them are obvious. The last two  are  discussed below. 

\begin{figure}[!ht]
\leavevmode
\begin{center}
\setlength{\unitlength}{1mm}
\begin{picture}(160,65)
\put(0,0){\makebox(0,0)[lb]{
\epsfig{file=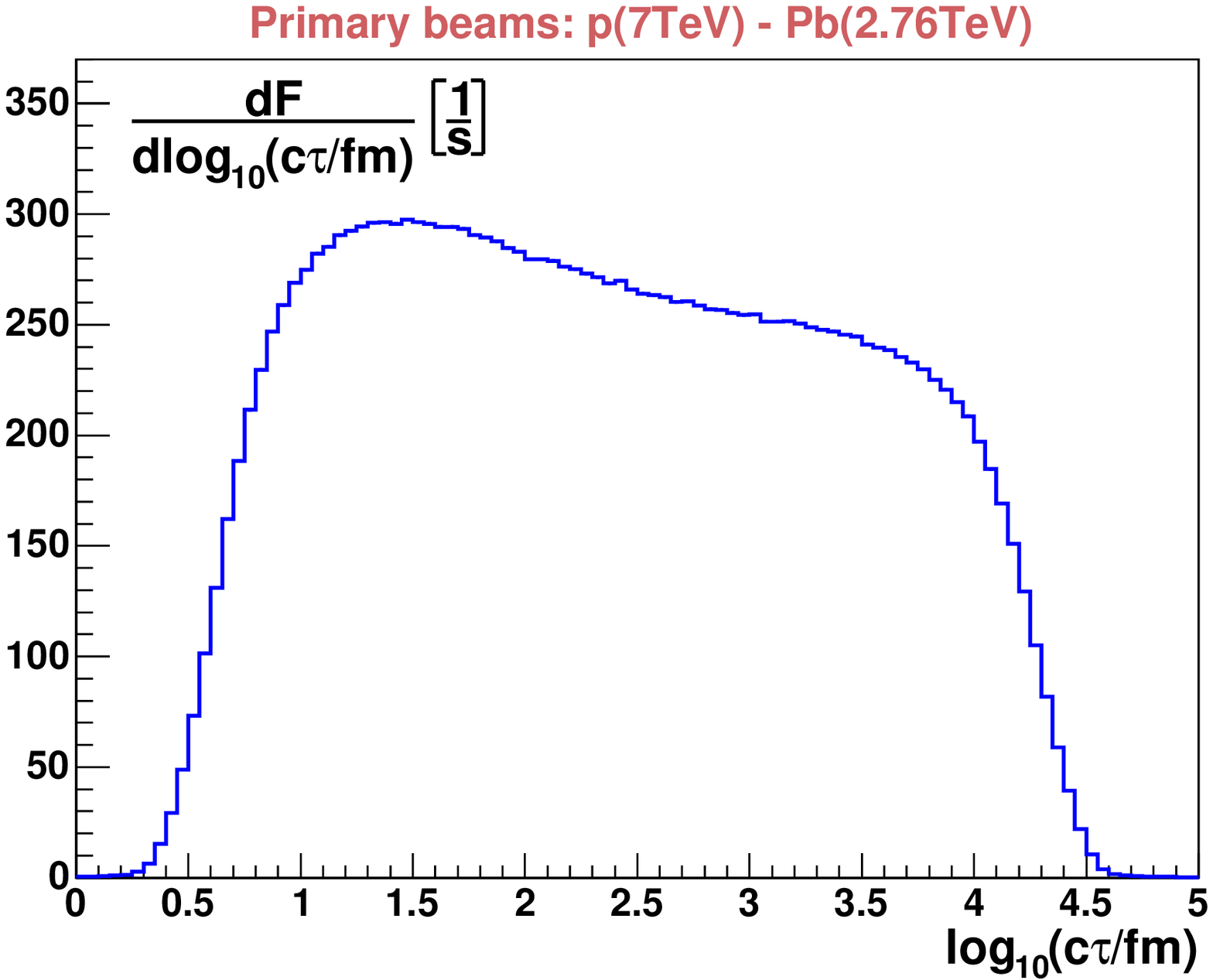, width=80mm,height=70mm}
}}
\put(75,0){\makebox(0,0)[lb]{
\epsfig{file=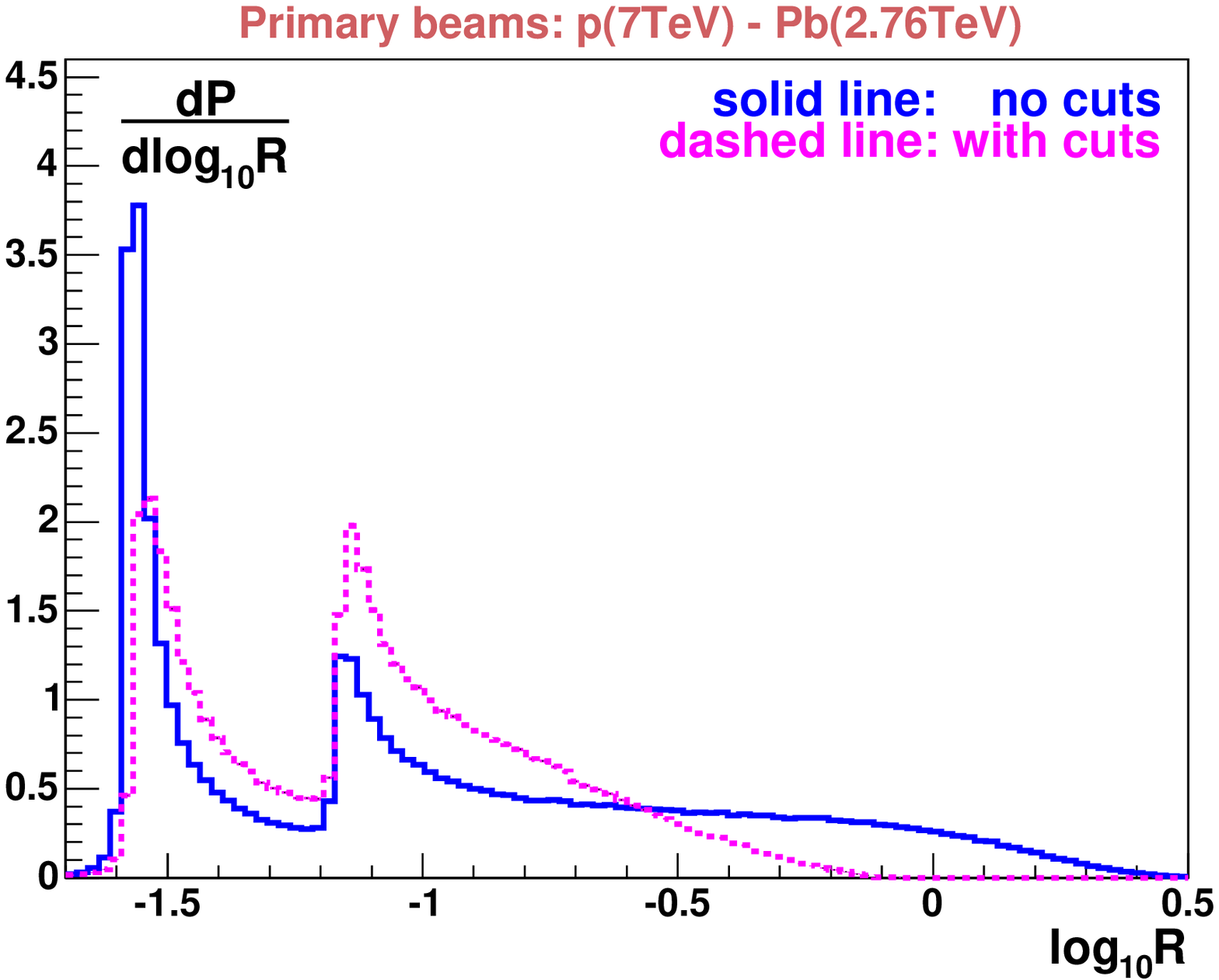, width=80mm,height=70mm}
}}
\put( 40,-5){\makebox(0,0)[cb]{\bf (a)}}
\put(115,-5){\makebox(0,0)[cb]{\bf (b)}}
\end{picture}
\end{center}
\caption{\sf The distributions of: (a) the $W$-boson lifetime and (b) $\log_{10} R$.
         See the text for more details.}
\label{fig:lifetime}
\end{figure}
In Fig.~\ref{fig:lifetime}a we show the distribution of the logarithm of the total 
path-length, $c \tau$, of $W$-bosons produced in the $p$--$Pb$ collisions, 
measured  in the rest frame 
of the $Pb$-bunches%
   \footnote{We postpone the discussion of Monte Carlo generator, which was 
             used to produce this plot,   till the next section.}.
Before decaying,  the $W$-bosons  formed in the collisions of the fastest 
(mainly valence) quarks of the incoming nucleon,  travel the distance
of $\sim 10^4$~fm,  while those formed in collisions of  the slowest 
(mainly sea) quarks of the 
incoming nucleon travel the distance of $\sim 10$~fm.

The $W$-bosons are produced in the process of Drell--Yan-like annihilation of 
quarks and antiquarks%
  \footnote{Formally, the partonic picture of hadrons
            is valid in the infinite-momentum frame, e.g.\ in the rest frame of 
            the LHC experiments. In this frame the  $W$-boson formation 
            can be pictured as the annihilation of the $SU(2)_L$-partner partons. 
            In the nucleus rest frame the same process can be pictured as the 
            bremsstrahlung of the $W$-boson from the incoming nucleon's quark. 
            The results and conclusions  presented in this 
            paper do not depend upon the choice of the reference frame.}.
The quantum-uncertainty of the longitudinal position $z$ of the quark--antiquark annihilation 
process with respect to the 
position of the nucleus is determined by the momentum of the quark (antiquark) 
which has been removed from the nucleus. 
In the rest frame of the nucleus it  is uncertain 
within the Ioffe-length \cite{Ioffe}:  
\begin{equation}
L_{\text{Ioffe}}(x_A) = \frac{1}{2 M_A x_A}\,, 
\label{Ioffe}
\end{equation}
where $M_A$ is the mass of the nucleus and 
$x_A= p_{\text{parton}}/p_{\text{nucleus}}$
is the infinite-momentum-frame fraction of the nucleus momentum carried by 
the parton taking part in formation of  the $W$-boson. Note,  that 
for those of partons which can be associated to the 
individual nucleons of the nucleus: 
$x_A= p_{\text{parton}}/(A\,p_{\text{nucleon}}) = x_N/A$,
if the effects of the nucleon Fermi motion is  neglected. 
The annihilation process can thus be localized, within the volume of the individual nucleons 
of the nucleus if $x_A \sim  M_N/M_A$, or, at the LHC energies, 
up  to the distances of  
$ L_{\text{Ioffe}}  \sim 10^3$~fm, 
if the annihilation process involves the  quark (antiquark)  carrying 
the infinite-momentum-frame fraction 
$x_A \sim 10^{-4} M_N / M_A$ of the nucleus momentum.

In order to show  that the LHC energy is sufficiently high for the $W$-boson decay 
length to be  larger than 
the total distance  between the position $r$ of the 
quark--antiquark annihilation  and the position  where the $W$-boson  exits 
the nuclear medium we construct, for each annihilation event,  
the ratio $R$ defined as:  
\begin{equation}
R=\frac{\langle l_A \rangle + L_{\text{Ioffe}} }{c \tau_W \gamma_W}.
\label{ratioR}
\end{equation}
where $\langle l_A\rangle$ is the average path-length of the $W$-boson 
in hadronic matter for the $W$-bosons created within the volume of the 
nucleus%
  \footnote{The calculation of this quantity is presented in Section~\ref{sec:targets}.}.
In order to satisfy the discussed requirements, this ratio must be $< 1$, 
both for the large momentum $W$-bosons, for which  
$\langle l_A\rangle \ll L_{\text{Ioffe}}$, and for the 
small momentum $W$-bosons, for which  $\langle l_A\rangle \gg  L_{\text{Ioffe}}$.
In Fig.~\ref{fig:lifetime}b we show  the probability distribution of $\log_{10}R$ 
for the $W$-bosons produced 
at the LHC in the collisions of protons with the lead nuclei.
The solid line represents the full sample of $W$-bosons, while the dashed
line the sample which satisfies the canonical LHC trigger requirements
\cite{ATLAS-TDR} for selection of the $W$-boson events.
Both plots are normalized such that they represent the probability distributions.
It is evident that the trigger acceptance cut removes  the $R \geq 1$ tail 
of the distributions which is populated by  events in which the $W$-boson is 
produced by  the  valence quarks of the nucleus and the very slow
antiquarks  of the nucleon. 
At the LHC,  all registered $W$-bosons have  their decay length 
larger than the distance between their creation point and the point where 
they leave the hadronic matter%
  \footnote{The residual probability that the  
               $W$-bosons  produced in the nucleon--nucleus
               collisions decay   inside  the nucleus can  be
               controlled  {\bf experimentally}    by  studying  
               the photon radiation yield  in events in which the $W$-boson 
               decays into the electron and the neutrino.}.
It is worthwhile to note that,  if one considers only the leading 
order process of the $W$-boson production, and  if the $W$-boson  momentum satisfies
$\langle l_A\rangle  \ll L_{\text{Ioffe}}$, then  $R= \Gamma_W/M_W$ for massless quarks,
where $M_W$ is the $W$-boson mass and $\Gamma_W$ is its total decay width.

In the leading process of the $W$-boson formation,  the 
longitudinal size of the  $W$-boson formation cell%
\footnote{The $W$-bosons are created quasi instantaneously.  
          In the rest frame of the $W$-boson  the size of the $W$-boson 
          formation  cell that satisfies the Heisenberg uncertainty principle 
          is:  $\delta r \sim 1/M_W$, 
          where  the $W$-boson mass $M_W$ is expressed in the units in which 
          $c=1$ and the Planck constant $\hbar=1$.},    
$\delta z = \gamma_W/M_W$, is  equal to the quantum uncertainty  
of the position of the nucleus with respect to the $W$-boson creation point, 
irrespectively of the $W$-boson momentum.
The nuclear-rest-frame picture of the process is the following one.
The quark (antiquark) of the incoming nucleon emits a virtual $W$-boson. 
The quark(antiquark)--virtual-$W$-boson  pair travels until the quark (antiquark) 
is absorbed within  the nucleus. Slow quarks are absorbed at the nucleus 
surface while the  fast ones are absorbed within the volume of one
of the nucleon of the nucleus. The $W$-boson becomes a free, on-shell 
particle as soon as its companion quark is absorbed. 
In  subsequent 
collisions the $W$-bosons can be considered as free, on-shell particles%
  \footnote{The {\bf experimental} procedure to 
            demonstrate that the $W$-bosons were formed before colliding  
            in nuclear matter would  be straightforward  if the remnants of the nucleus and of 
            the nucleon were measured. Such  a procedure would then be  
            equivalent to that used in the analysis of tagged 
            photoproduction events at the HERA collider. At the LHC,  
            the off-shellness of the $W$-bosons  will be partially controlled by 
            studying the topology of the particle flow 
            in the analyzed sample of events with the final $W$-boson signature.}.

Two examples  illustrating the space time picture of the $W$-formation and decay 
process at the LHC energies are given below.
For $W$-bosons produced at the lowest  trigger-accepted $\gamma_W \sim 10^2$,  
the nuclear-rest-frame size of 
their  formation cell of the order of  $0.25$~fm. 
These $W$-bosons are produced mainly by the valence 
quarks of the nucleus, i.e.\ they are fully formed within the volume of a nucleon
within a nucleus. For subsequent interactions with the spectator nucleons 
they  are free on-shell particles. They  travel the distance 
of $ \sim 10$~fm before decaying. 
For the $W$-bosons produced at  the highest trigger-accepted 
$\gamma_W= 10^5$ the nuclear-rest-frame size of the their  formation cell 
is by two orders of magnitude greater than the nucleus size. 
These $W$-bosons  are produced by the fastest quarks of the incoming 
nucleon by bremsstrahlung of the $W$-boson at the distance of  $ \sim 10^3$ fm from the nucleus centre. They 
become fully formed on-shell particles  at the nucleus surface where 
its companion quark (antiquark) is absorbed. Their decay length is $ \sim 10^4$~fm.

\subsubsection{Amplitudes and probabilities}
\label{subsubsec:amplprob}

The $W$-boson propagation amplitude  ${\cal M}_p$ could, 
in principle,  be modeled and  implemented 
in dedicated Monte Carlo programs. One  could then try to unfold 
its  selected  properties  from the analysis of the  
measured observables of the nucleon-nucleus collisions.
The goal of the femto-experiment is, however, to directly measure  
the $W$--nucleon collision observables.
These observables  must thus be expressed only in terms of the  squares of the 
amplitudes and must not depend upon the variables which cannot be measured.
In order to show that such measurements are feasible, 
we have to integrate the matrix element  ${\cal M}$  over  
each of  the variables that could never be determined  experimentally 
(whatever experimental set-up is proposed),  and to calculate its square.  

In the process shown in Fig.~\ref{fig:factorization} 
the momentum of the produced $W$-boson could in principle be determined 
by measuring the remnants of the nucleon  and of the nucleus and by 
measuring the energy flow of the $W$-recoil jet%
  \footnote{The ambiguity of assigning  a fraction of particles produced
            in the inelastic collisions of $W$-bosons to the remnants,  
            and/or to the recoil jet  could,
             in principle, lead to residual interference effects for $W$-bosons 
             produced at very small transverse momenta.
             This effect is neglected in this paper.
             Note,  that the $W$-boson recoil particles could, in principle,
             be fully resolved  from the particles produced in the 
             $W$-boson collisions owing to the lack of the colour interconnection 
             between these two systems.}. 
The  momentum of the outgoing $W$-boson 
could be determined  by measuring the momenta of outgoing 
leptons (even if measuring the neutrino momentum would be highly unpractical).

The remaining integrals are thus those over the 
unobserved position of the $W$-boson creation point,  $(x,y,z )$:
\begin{equation}
\begin{aligned}
 |{\cal M}_{int}|^2 
&
 = \int{dx} \int {dy} \int {dz} \; |{\cal M} (x,y,z)|^2  \\
& =  \int{dx} \int {dy} \int {dz} \; |\, {\cal M} _f(x,y,z)\; 
     {\cal M}_p \left( l_A(x,y,z)\right) \; 
     {\cal M} _d\,|^2,
\end{aligned}
\label{eq:master1}
\end{equation}
where we have left over, for simplicity,  the  dependence  of the amplitudes upon 
the kinematic variables.

The above expression can  be simplified. As we have discussed in the previous 
section,  each  of the $W$-bosons produced in the primary collision,
irrespective  of its   momentum,  is  ``fully formed''  
before arriving at the position of the target.
In addition, as we shall discuss in detail in Section~\ref{sec:targets}, 
its path-length in hadronic matter is independent of the exact 
formation-point position within the $W$-boson production cell.
Therefore the  $(x,y,z)$-dependence of the relative phases of the  ${\cal M} _f$ 
and   ${\cal M}_p$ amplitudes
can be neglected and the position of the $W$-boson creation point can be replaced by the 
``coarser granularity'' position of the $W$-boson  formation cell:  $(x_p,y_p,z_p)$.
 Consequently  the formula~(\ref{eq:master1}) can be written in the following form:
\begin{equation}
 |{\cal M}_{int}|^2 =  |{\cal M} _d|^2 \;   \int{dx_p} \int {dy_p} \int {dz_p}\; 
 |{\cal M}_f(x_p,y_p,z_p) |^2\;  
 \left|{\cal M}_p\left(l_A(x_p,y_p,z_p)\right)\right|^2\,.
\label{eq:master2}
\end{equation}

We have,  so far,  demonstrated  that the partonic collisions producing  the $W$-bosons  
can be considered, in the studies of  the $W$-boson  propagation in the nuclear medium,  
as a probabilistic random emission source 
producing the $W$-boson beam. The only difference of such a beam with 
respect to the $W$-beam of the `gedanken experiment' is that its source
is randomly distributed in space within the nuclear volume and, 
for the high-momentum $W$-bosons,  outside the nuclear volume.

\subsubsection{Cross sections}
\label{subsec:xsecs}

The factorized form of the integral~(\ref{eq:master2}) allows us to  directly link 
the  rate of of those of the nucleon--nucleus
collision events  shown in Fig.~\ref{fig:factorization},  in which the produced $W$-boson 
collided in the nucleus,   
to the  the $W$--nucleon collision  inclusive cross section.  
Indeed, the factorization of the squared  amplitudes allows us to express 
the  rate ${\cal N}_{evt}$ in the following form: 
\begin{equation}
\begin{aligned}
{\cal N}_{evt} (p_n, p_A, p_W^{in}, p_W^{out}, p_1 | A)  =  
\int{dx_p} \int {dy_p} \int {dz_p} \;     
{\cal N}_{W} ( p_n,p_A,p_W^{in}, x_p, y_p, z_p  | A) 
& \\
\times {\cal \sigma}_{Wn}^{incl} ( p_W^{in} , p_W^{out})    
\; l_A( x_p,y_p,z_p,  p_W^{in},p_A  )\; \rho_A(p_A) 
\; {\cal P}_{decay}( p_1, p_W^{out} - p_1) & 
\end{aligned}
\label{eq:master3}
\end{equation}
where
$  {\cal N} _{W} ( p_n,p_A,p_W^{in}, x_p, y_p, z_p  | A)  $
is the unit volume-rate of the $W$-bosons produced at the cell-position 
$ (x_p, y_p, z_p) $,     
$ {\cal \sigma}_{Wn}^{incl} ( p_W^{in} , p_W^{out})  $ 
is the $W$--nucleon collision inclusive cross section, 
$  l_A( x_p,y_p,z_p,  p_W^{in},p_A)   $ is the path-length 
of the $W$-boson in the nucleus characterized by  
its atomic number $A$ and  the nuclear  density%
  \footnote{Note,  that the path-lengths and the  nuclear density must be 
            calculated in the reference frame in which the kinematic 
            variables are determined.}
$\rho _A(p_A)$, 
and  ${\cal P}_{decay}( p_1, p_W^{out} - p_1)$
is the probability of the decay of the $W$-boson having the momentum $ p_W^{out}$
into the  $SU(2)_L$-doublet leptons  
of the four-momenta $p_1$ and $  p_W^{out} - p_1  $. 
The above  formula is valid for the $W$--nucleon cross section 
$\sigma_{Wn} ~\ll~ 1$~fm$^2$. This condition assures that 
the $W$-boson interacts at most once within the nuclear volume.

The result of the integration is:
\begin{equation}
\begin{aligned}
{\cal N}_{evt} (p_n, p_A, p_W^{in}, p_W^{out}, p_1 | A) = \;
&  
{\cal F}_{W}  ( p_n,p_A,p_W^{in}  | A)   \\
\times {\cal \sigma}_{Wn}^{incl} ( p_W^{in} , p_W^{out}) \;  
& 
\langle l_A(  p_W^{in}, p_A) \rangle \;
\rho _A (p_A) \; 
{\cal P} _{decay} ( p_1, p_W^{out}-p_1   ),
\end{aligned} 
\label{eq:master4}
\end{equation}
where ${\cal F}_{W} ( p_n,p_A,p_W^{in} | A)$ is the total flux of the $W$-bosons, 
and  
\begin{equation}
\langle l_A( p^W _{in},p_A )\rangle  = \int{dx_p} \int {dy_p} \int {dz_p} \;     
{\cal P}_{W}(p_n, p_A, x_p,y_p,z_p,p^W _{in} | A) 
\; l_A( x_p,y_p,z_p, p^W _{in},p_A )
\label{eq:master5}
\end{equation} 
is the average path-length of the $W$-boson in hadronic matter.
In this formula  ${\cal P}_W$ is  the probability density to  produce  
the $W$-boson carrying the momentum $p^W _{in}$
within the formation  cell  positioned at  $r_p = (x_p,y_p,z_p)$. 
It is formally expressed as:    
\begin{equation}
 {\cal P} _{W}(p_n, p_A, x_p,y_p,z_p,p^W _{in} | A) =     
 \frac{ {\cal N}_{W}(p_n, p_A, x_p,y_p,z_p,p^W _{in} | A)}  
      { \int{dx_p} \int {dy_p} \int {dz_p}\;   
        {\cal N} _{W}(p_n, p_A, x_p, y_p, z_p, p^W_{in} | A) }\,. 
\label{eq:master6}
\end{equation} 
 
The formula~(\ref{eq:master5}) becomes  particularly simple if 
explicitly integrated over the outgoing lepton momentum  $ p_1$.
Since the integrated  decay probability 
is equal to $1$, the lepton-momentum integrated rate can be expressed 
as%
\footnote {The reason why we have  kept 
           the decay-topology dependent probabilities will become clear in the 
           following section devoted to the discussion of the framework for the 
           collisions of polarized $W$-boson beams.}:
\begin{equation}
{\cal N}_{evt}(p_n,p_A , p^W _{in},p^W _{out} |A) 
=    {\cal F}_{W}(p_n, p_A, p^W _{in} | A)  
 \; {\cal \sigma}_{Wn}^{incl} (p^W _{in},  p^W_{out}) 
 \; \langle l_A( p^W _{in},p_A) \rangle  \;  \rho _A(p_A)\,.
\label{eq:master7}
\end{equation}
   
The above formula, if written in the Lorentz-reference frame
in which the nucleus is at rest 
and the $W$-boson has only the longitudinal component of 
its momentum:  
$p_W^{in} = (0,0, \sqrt{[s_{Wn} - (M_W + M_n)^2][s_{Wn} - (M_W - M_n)^2]}/2M_n, 
             ( s_{Wn}- M_W^2 - M_n^2)/2M_n)$,
where  $s_{Wn}$ is the CMS energy of the $W$--nucleons
collisions, is equivalent to the `gedanken experiment' formula~(\ref{gedanken}),
summed up over the spin indices. This reference frame will be called hereafter
the {\it collinear $W$--nucleon collision} frame. 

We have thus demonstrated that the femto-experiment analysis framework 
is identical to that of the `gedanken experiment'. 
In order to measure $s_{Wn}$
and the   $p^W_{out}$ dependence of the inclusive cross section  
${\cal \sigma}_{Wn}^{incl} (p^W _{in}, p^W_{out})$,
one  has  to measure the event rate ${\cal N}_{evt}$, 
the total flux of $W$-bosons   ${\cal F}_{W}(p_n, p_A, p^W _{in} | A) $,   
and calculate the average target lengths  $\langle l_A( p^W _{in},p_A) \rangle $. 

\subsubsection{Framework for polarized $W$-beam}
\label{subsec:frampolar}

In the previous sections we have proposed the framework to analyze 
the  nucleon--nucleus collision   observables 
in terms of the $W$--nucleon collision ones  for the collisions 
of the unpolarized $W$-bosons assuming that 
the  identification of
the spin state of the final $W$-boson cannot be  made. In this  section we generalize
this framework to describe the inclusive scattering of polarized $W$-bosons.
Such a framework, as we shall demonstrate in the 
following sections, and in the forthcoming paper \cite{forthcoming}, 
will be very useful 
in exploring  spin asymmetries in the collisions
of the $W$-bosons with hadronic matter -- in particular in measuring  the 
$W$--nucleon collision-energy dependence of the inclusive cross-section asymmetry 
for longitudinally and  and transversely polarized $W$-bosons.
The key point,  which has triggered our interest 
in femto-experimenting at the LHC  is 
that we can propose a method to tune the  
polarization of the $W$-boson beam and to measure the polarization of the 
outgoing $W$-bosons.

In Section~\ref{subsubsec:amplprob}  we have demonstrated  
that the partonic collisions producing  $W$-bosons  
can be treated as a probabilistic random emission source 
producing the $W$-boson beam. 
For the incoherent source producing the $W$-bosons  
with the polarization $\lambda_{in}$,  the number of events 
 ${\cal N}^{\lambda_{in}} _{evt} (p_n, p_A, p_W^{in}, p_W^{out}, p_1 | A)$
in which the $W$-boson produced with momentum $p_W^{in}$ 
collides and changes its momentum to $p_W^{out }$  
can be expressed by the following formula:
\begin{equation}
\begin{aligned}
{\cal N}^{\lambda_{in}} _{evt} (p_n, p_A, p_W^{in}, p_W^{out}, p_1 | A) 
=    {\cal F}^{\lambda_{in}} _{W}  ( p_n,p_A,p_W^{in}  | A)   
& \\
 \times \left| \sum _{\lambda_{out} } 
 {\cal S}^{\lambda_{in}, \lambda_{out}}_{Wn} ( p_W^{in} , p_W^{out}) \;  
 {\cal D}^{\lambda_{out}}_{W}( p_1, p_W^{out}-p_1 ) \, \right|^2\;
&
 \langle l_A(  p_W^{in}, p_A)   \rangle  \;  \rho _A(p_A) \,,
\end{aligned} 
\label{eq:masterspin}
\end{equation}
where $  {\cal F}^{\lambda_{in}} _{W}  ( p_n,p_A,p_W^{in}  | A) $ is the flux 
of the $W$-bosons with the polarization $\lambda_{in}$, 
${\cal S}^{\lambda_{in}, \lambda_{out}}_{Wn} ( p_W^{in} , p_W^{out})$
is the amplitude  of the $W$-boson scattering on a nucleon,
leading to a  change of its momentum to   $ p_W^{out} $ 
and to a change of its polarization to  $\lambda_{out}$,
and ${\cal D}^{\lambda_{out}}_{W}( p_1, p_W^{out}-p_1 )$ 
is the decay amplitude of the $W$-boson into two $SU(2)_L$-doublet fermions.

The important property  of this  formula reflects  the fact that  
the spin of the outgoing $W$-boson could not  be directly measured, 
even if we were able 
to measure all particles produced in the nucleon--nucleus collisions. 
Therefore,  a coherent summation over the pure spin states of the outgoing $W$-boson 
has to be made.

The relation of spin-dependent amplitudes to the $W$--nucleon inclusive cross 
section can be written as follows:
\begin{equation}
 \int d^3p_1 \; 
 \frac{\sum_{\lambda_{in}} {\cal F}^{\lambda_{in}}_{W} }{  {\cal F} }\;   
 \frac{| \sum _{\lambda_{out} } {\cal S}^{\lambda_{in}, \lambda_{out}}_{Wn} 
         {\cal D}^{\lambda_{out}}_{W})|^2 }{  {\cal \sigma}_{Wn}^{incl} }  = 1\,.
\label{eq:masterspin1}
\end{equation}
This relation defines the units  of the global normalization factor of ${\cal S}$  
to be the cross-section units.    

The observables describing the inclusive collisions of the polarized $W$-bosons:  
$W^{\lambda_{in}} + n \rightarrow  W^{\lambda_{out}} + X$
have to be unfolded from the observed event rate. 
The most convenient reference frame in which such unfolding can be made 
is the $W$-boson rest frame in which 
the spin quantization axis is chosen to be the  
{\it collinear $W$--nucleon collision} frame $z$-axis.
The spin dependent observables can be  
unfolded from the  measured  angular 
distribution of the $W$-boson  decay products. 

In the  $W$-boson rest frame the number of events in which 
the  charged lepton is emitted at the angles $(\theta,\phi)$,  with respect to the 
spin quantization axis,  can be expressed as follows:
\begin{equation}
\begin{aligned}
{\cal N}^{\lambda_{in}}_{evt}(p_n, p_A, p_W^{in}, p_W^{out}=0, \cos\theta,\phi\,| A)
=
\hspace{62mm}
& \\
  \frac {3}{4\pi} \sum_{\mu,\nu} \sum_{\lambda_{1},\lambda_{2}}
   {\cal F}^{\lambda_{in}}_{W}  ( p_n,p_A,p_W^{in}  | A) \;
{\cal S}^{\lambda_{in}, \lambda_{1}}_{Wn} ( p_W^{in}, p_W^{out})\;
{\cal S}^{*\lambda_{in}, \lambda_{2}}_{Wn} ( p_W^{in}, p_W^{out})\;
\left|{\cal T}^{\mu \nu}\right|^2 \; 
\hspace{1mm}
& \\
 \times
  D^{1*}_{\lambda_{1} (\mu -\nu)} (\cos\theta,\phi) \;
  D^{1}_{\lambda_{2} (\mu-\nu)}    (\cos\theta,\phi)\;  
 \langle l_A(  p_W^{in}, p_A)   \rangle \;  \rho _A (  p_A ) \,, 
\hspace{5mm}
&
\
\end{aligned} 
\label{eq:masterspin3}
\end{equation}
where  
${\cal N}^{\lambda_{in}}_{evt}$ is the total number of events having the kinematic 
variables $(p_n, p_A, p_W^{in}, \cos\theta,\phi)$, 
${\cal T}^{\mu \nu}$ are the Standard Model helicity amplitudes of the decay of the 
$W$-boson into leptons having, respectively, the spins $\mu$ and $ \nu$, 
$D^{1}_{\lambda_{2} (\mu-\nu)}    (\cos\theta,\phi)$ and 
$ D^{1*}_{\lambda_{1} (\mu -\nu)}(\cos\theta,\phi)$ are the 
matrices corresponding to Wigner rotations for spin-$1$ particles, 
and  $\rho _A (  p_A)$ is the nucleus density in the $W$-rest frame.

The above formula is derived using the helicity amplitude formalism presented 
in Ref.~\cite{Jackson}
and expressing the spin density matrix of the outgoing $W$-boson as:
\begin{equation}
{\cal \rho}^{\lambda_{1}, \lambda_{2}}_{W_{out}}=
\sum_{\lambda_{in} ,\lambda }   {\cal \rho}^{\lambda_{in} \lambda}_{W} \;
{\cal S}^{\lambda_{in}, \lambda_{1}}_{Wn} ( p_W^{in}, p_W^{out})\;
{\cal S}^{*\lambda, \lambda_{2}}_{Wn} ( p_W^{in}, p_W^{out})\;
\label{eq:masterspin4}
\end{equation}
 where
\begin{equation}
{\cal \rho}^{\lambda_{in} \lambda}_{W_{in}}= \delta^{\lambda_{in} \lambda} 
\; \frac{ {\cal F}^{\lambda_{in}} _{W} }{ 
          \sum_{\lambda_{in}} {\cal F}^{\lambda_{in}}}\,.
\label{eq:masterspin5}
\end{equation} 

The full information on the amplitudes (including phases) 
of the polarized $W$-scattering is contained in  the 
${\cal S}^{\lambda_{1}, \lambda_{2}}_{Wn}$ matrix elements.
In order to measure their   $s_{Wn}$ and $p^W_{out}$ dependence, we need 
to polarize the beam of $W$-bosons,
and to  measure their flux ${\cal F}^{\lambda_{in}} (p_n, p_A, p^W _{in} | A) $  
and the event rate 
${\cal N} _{evt} (p_n, p_A, p_W^{in}, p_W^{out}=0, \cos\theta,\phi\, | A)$.

\subsection{Measuring kinematic variables at LHC}
\label{subsec:kinvar}

If the beam of the $W$-bosons at the LHC 
could be momentum and spin tagged, and if the
momentum  and spin of the outgoing $W$-boson could be
unambiguously determined from the 
angular distribution of the final-state leptons, then 
the CMS-energy and polarization dependence of the $W$--nucleon cross section
(polarized matrix elements) could  be directly determined from the observed 
event rates using the above formulae%
  \footnote{The procedure of 
            determining the inclusive polarized $W$--nucleon cross section  
            would be analogous to the determination of the polarized 
            $\rho$--nucleus cross section using 
            $e$--$A$ collisions  of polarized electron beam and  
            tagging  the initial and the final-state electron momenta 
            and the momenta of the pions produced in the $\rho$-meson decays.}. 
However, at  the LHC the remnants of the nucleon and of the nucleus cannot be 
measured%
  \footnote{The highest energy circular collider where
            a 4$\pi$-detector, integrated with the machine lattice,  
            was  designed is the RHIC 
            collider~\cite{Krasny-Yale, Krasny-Snowmass, Krasny-Trento}.}.
Similarly, the momentum of the outgoing neutrino cannot be directly measured.
As consequence the projected momenta  of the incoming 
and of the outgoing $W$-boson on the nucleon--nucleus collision axis 
cannot be measured. The projected momenta of the 
incoming and of the outgoing $W$-bosons on the plane perpendicular    
this axis can be measured only if the final state particles  can be 
associated either to the recoil system produced in the primary collision
producing the $W$-boson,  or to the group of particles produced
in the subsequent collision of the $W$-boson. Note  that the above two 
systems fragment independently and 
do not have any colour interconnection (their correlation
and quantum interferences would be important 
only if the nuclear de-excitation global variables were measured). 
These issues  will be discussed in more detail in  Ref. \cite{forthcoming} -- in the 
following,  we shall assume that an algorithm for such an association 
can be devised and that one  can provide,  on the event-by-event basis,  
the estimators of the transverse momenta of the two systems, 
$\vec{p}_T^{\;\text{recoil}}$ and  $\vec{p}_T^{\;\text{subs}}$. 
The transverse momentum of the outgoing $W$-boson 
can thus be expressed as  
$\vec{p}_T^{\;W, out} = -(\vec{p}_T^{\;\text{subs}} + \vec{p}_T^{\;\text{recoil}})$,
and the transverse momentum of the neutrino as 
$\vec{p}_T^{\;\nu} = \vec{p}_T^{\;W, out} - \vec{p}_T^{\;l}$.

\subsection{Polarization of $W$-bosons at LHC}  

The spin of the $W$-boson beam particles cannot be measured  on the event-by-event basis.
The $W$-beam can be considered as  an incoherent mixture 
of the three beams representing the three polarization states of the $W$-bosons.
The spin-dependent observables
must thus be defined in terms of the sums of the cross sections weighted
by the relative intensities of the beams with the three polarization states.
The polarization of the outgoing $W$-boson is encoded into the angular distributions
of its decay  products. Since the longitudinal momentum of
the outgoing $W$-boson cannot be measured at the LHC, 
the two  spin-analysis angles are reduced to only one, 
chosen in this paper  as the angle,  $\phi_l^t$,  between of the produced charged lepton 
and the outgoing $W$-boson in the plane perpendicular to the  
nucleon--nucleus collision axis.

\subsection{Unfolding of $W$--nucleon collision observables in LHC environment}
\label{subsec:unfolding}

At the LHC,  the $W$--nucleon collision observables 
will have to be unfolded from the observed rates of events containing 
the final-state charged lepton (preferably electron or muon)  
and missing transverse energy. 
These events will be characterized by: 
the reconstructed lepton momentum $ p_l$, 
the reconstructed transverse momentum of the neutrino $\vec{p}_T^{\;\nu}$, 
and the reconstructed transverse momentum of the particle system associated
with the $W$-boson recoil system%
\footnote{Other  topological variables
          characterizing the hadronic and leptonic energy flow 
          in selected events could be very useful 
          for detailed studies of  the mechanism of the $W$-boson interactions.
          This aspect is not discussed in this paper.}
$\vec{p}_T^{\;\text{recoil}}$.
 
The relationship between the rates of events containing  the above leptonic signatures
and the $W$--nucleon collision observables
can be written as:
\begin{equation}
\begin{aligned}
{\cal N}_{\text{LHC}}(p_n,p_A, p_T^{\text{recoil}},  p_T^{\nu}, p_l\, |A) =
\hspace{80mm}
& \\  
 \int d^3p_W^{in} \int d^3 p_W^{out}\; 
  \delta^{(2)}\left(p_T^{W,\,in} -  p_T^{\text{recoil}}\right)  
    \delta^{(2)}\left(p_T^{\nu} + p_T^{l} - p_T^{W,\, out}\right) 
\hspace{20mm}
& \\ 
 \times {\cal F}_{W} ( p_n,p_A,p_W^{in}\,| A) 
 \left[1 - \sigma_{tot}^{abs}( p^W _{in})\; 
  \langle l_A( p^W _{in}, p_A  )\rangle\;   \rho _A(p_A) \right] 
& \\
\times \bigg[ 1- \sigma_{tot}^{incl}( p^W_{in} )\;  
    \left\{1 - \Theta\left(|p^W _{in} - p^W _{out}| - \epsilon\right)\right\}
 \; \langle l_A( p^W _{in}, p_A  )\rangle \;  \rho _A(p_A)\;\;   
& \\
 +\; {\cal \sigma}_{Wn}^{incl}(p^W _{in},  p^W_{out})\; 
  \Theta\left(|p^W _{in} - p^W _{out}| - \epsilon\right)  
   \; \langle l_A( p^W _{in}, p_A  )\rangle \;  \rho _A(p_A)\bigg]
& \\
  \times\; {\cal P}_{\text{decay}}( p^W _{out}, p_l)
& \,,
\end{aligned}
\label{eq:general}
\end{equation}
for the measurement of the total inclusive cross section, and:
\begin{equation}
\begin{aligned}
{\cal N}^{pol}_{\text{LHC}}(p_n, p_A, p_T^{\text{recoil}}, p_T^{\nu}, p_T^l, 
                        |p_l|, \phi^t_l )\, | A) = 
\hspace{60mm}
&\\
\frac {3}{4 \pi} \sum_{\mu, \nu} \sum_{\lambda_{in}, \lambda_{1}, \lambda_{2} } 
 \int d^3 p_W^{in} \int d^3 p_W^{out}  
 \delta^{(2)}(p_T^{W,\,in} -  p_T^{\text{recoil}})  
 \delta^{(2)}(p_T^{\nu} + p_T^{l} - p_T^{W,\,out})
& \\
 \times {\cal F}^{\lambda_{in}} _{W}  ( p_n,p_A,p_W^{in}\,  | A)  
  \left[ 1 - \sigma_{tot}^{abs}( p^W _{in})\; 
  \langle l_A( p^W _{in}, p_A  )\rangle \;  \rho _A(p_A) \right] 
&\\
 \times 
{\cal S}^{\lambda_{in}, \lambda_{1}}_{Wn} ( p_W^{in}, p_W^{out})\;
{\cal S}^{*\lambda_{in}, \lambda_{2}}_{Wn} ( p_W^{in}, p_W^{out})\; 
\langle l_A( p^W _{in}, p_A)\rangle \;  \rho _A(p_A)
& \\
\times |{\cal T}^{\mu \nu}|^2 \; 
D^{1*}_{\lambda_{1} (\mu -\nu)} (\cos\theta,\phi) \;  
D^{1}_{\lambda_{2} (\mu-\nu)}(\cos\theta,\phi)
& \,
\end{aligned} 
\label{eq:general1}
\end{equation}
for the measurement of the spin dependent matrix elements%
  \footnote{The angles $\theta$ and $\phi$ in eq.~(\ref{eq:general1})
             are expressed in terms of  $ p_W^{out}$, $p_T^l$, $|p_l|$ 
             and $\phi^t_l$.}.
$\sigma_{tot}^{incl}( p^W _{in})$  is the total inclusive cross section 
and the $\sigma_{tot}^{abs}( p^W _{in})$ is the $W$-boson absorption 
cross section. 
The symbol  $\delta^{(2)}(x)$ represents the $2$-dimensional Dirac 
$\delta$-function,  and $\Theta(x)$ represents the step function: 
$ \Theta(x) = 0 $ for $x < 0$,  and $\Theta(x) = 1 $ for $x \geq 0$.
All $p_T$s in the above formulae are $2$-dimensional vectors in the
transverse momentum plane, while other $p$s denote three-momenta.
The introduction of the $\Theta(x)$  function and its parameter $\epsilon$ 
is technical. It allows to define the  
two origins of the observed events: 
those in which the  outgoing $W$-bosons interacted in the hadronic matter 
and those in which they did not. 
Such a separation is not physical. The  measurement 
results must not depend upon the way how these subsamples are defined.
Indeed, the $\epsilon$ dependence of the sum of the two corresponding 
terms vanishes, in any  concrete unfolding procedure, as soon as  
the    $\epsilon$ value   is chosen to be smaller than the resolution with which  
the sum of the transverse momenta,  
$\vec{p}_T^{\;\text{recoil}} + \vec{p}_T^{\;\nu} + \vec{p}_T^{\;l}$,
is measured%
  \footnote{For the spin-dependent formula the technical separation 
            is not necessary because  
            ${\cal S}^{\lambda_{1}, \lambda_{2}}$ contains both 
            contributions, whatever criteria of the 
            technical splitting  of the full sample of events into 
            the transmission and interaction subsamples are used.}. 
The term $[1 - \sigma_{tot}^{abs}(p^W_{in}) \langle l_A(p^W _{in}, p_A)\rangle
          \;\rho _A(p_A)]$ describes the disappearance of the $W$-bosons. 

The precision of unfolding of the $W$--nucleon collision observables 
will be driven by the statistical 
precision of measuring event rates and by the systematic precision 
of measuring the kinematic
variables. 
However,  the most important source of the  measurement errors  will
be the uncertainties in  the kernels of the integral equations. 
The quest for  their high precision is amplified 
not only by the fact that  only the integrated 
rates will be measured at the LHC,  but also  by the fact that in the 
Monte-Carlo-based unfolding methods, the  propagation of 
the kernel errors to the measured quantity cannot be fully controlled.

The kernels of the integral equation contain the expressions which 
can be precisely calculated and those which depend  
upon the external experimental input. The dominant 
contribution to a measurement error will be due to 
the uncertainty in the kinematic,  spin and nuclear  dependence of the flux of the 
$W$-bosons%
  \footnote{Note that, conceptually, the unfolding   methods for  the 
            $W$-beam fluxes  at the LHC will  be similar to those  
            developed in the past for the SPS neutrino fluxes.},
${\cal F}^{\lambda_{in}} _{W}  ( p_n,p_A,p_W^{in}  | A)$   
and the $W$-boson momentum dependence of the average target length.
The precision with which they could be presently calculated will 
likely turn out to be insufficient to explore the rare collision of $W$-bosons.
Therefore one has  to propose special  measurement procedures reducing 
the influence of these uncertainties on the measurement of 
the $W$--nucleon collisions observables. In the following section of this 
paper we shall propose such procedures fulfilling the third, 
necessary and  sufficient  condition,  which must be satisfied
for   ``configuring'' the nucleon--nucleus collisions at the LHC 
as the femto-experiment for the investigation of the 
high-energy $W$-bosons interactions with matter.

\section{Beams}
\label{sec:beams}

\subsection{Theoretical control} 
\label{subsec:techcontol}

The $W$-boson beams
will be generated,  at the Interaction 
Points (IPs) of the LHC collider,  by hard collisions of the 
constituents of the colliding particles: 
quarks, antiquarks and gluons. 
The leading process is the Drell--Yan-like annihilation of
quarks and antiquarks -- the members of $SU(2)_L$ doublets with opposite
values of the third weak-isospin component. Non-diagonal elements of the 
Cabbibo-Kobayashi-Maskawa (CKM) matrix lead to mixing between the 
quark generations. The $W$-bosons do not couple directly to gluons.
However,  gluons  can couple  to  a 
quark--antiquark pair,  providing  a supplementary source of 
quarks and antiquarks for the Drell--Yan-like process.   
This latter process, even if next-to-leading 
in the strong interaction coupling constant $\alpha_s$,
influences considerably the $W$ production rates at the LHC
since the high energies of the colliding beams allow to reach
the low Bjorken-$x$ region where the gluon density in the nucleon
becomes seizeable. 

The cross section of the $W$-boson production at hadronic colliders  is
calculated as a product of the hard scattering cross section  and  
parton distribution functions (PDFs). The hard scattering cross section
can be calculated using perturbative methods with the precision 
of  ${\cal O} (\alpha_s^2)$ at the high transverse momentum of the 
$W$-boson~\cite{W_QCD}. For the transverse momentum of the $W$-boson 
substantially smaller than its mass,  the fixed order
perturbative result has to be complemented by a re-summation of large 
logarithmic correction to all orders in $\alpha_s$~\cite{W_QCD1}.
  
The electroweak radiative cross sections modify the calculated 
cross section with terms proportional to $\alpha_{QED}$.
Contrary to the QCD corrections, they depend upon the 
flavour of produced leptons. The electroweak radiative
corrections are dominated by the virtual corrections 
to $W$-propagator and  vertices and by the QED radiation 
by  the final state lepton(s)~\cite{Baur:1998,Dittmaier:2002}. 
The photon radiation from the  quarks, the interference terms 
as well as higher-order radiative corrections, including 
re-summation of multiple soft-photon emissions, must also be 
included in  high-precision 
calculations~\cite{Baur:1998,Dittmaier:2002,WINHAC:2003}.

By the time when the LHC will start taking data
the $W$ production processes are expected to be controlled  
theoretically to $\sim 1$--$2\%$ for fixed PDFs.
The detailed discussion of  theoretical and experimental aspects of 
the $W$-boson production in hadronic colliders  can be found 
e.g.~in Ref.~\cite{Haywood:2000}, and recently in Ref.~\cite{Nadolsky}. 

The presently known QCD calculation methods do not allow the 
PDFs to be directly calculated. The perturbative methods allow, however,
to derive  the relevant PDFs from  the deep-inelastic scattering data  
and to extrapolate them to the virtualities scales involved in $W$-boson
production.
Uncertainties of the PDFs arising from various theoretical 
sources have been propagated to the uncertainties in the 
predicted $W$-production cross section in Ref.~\cite{Tung}.
The uncertainty of the PDFs will determine the precision 
of the predicted $W$-production cross sections at the LHC, 
and its rapidity and transverse-momentum dependence. 
The $W$-production cross section for the $p$--$p$ collision will likely be 
controlled to the $\sim 5\%$ precision at the start-up of the 
LHC operation.  This uncertainty will be gradually 
diminished with increasing global 
understanding of the LHC data.

The nuclear PDFs  suffer from large uncertainties, in particular
in the small-$x$ domain where only fixed-target deep-inelastic
scattering data, confined to low-virtuality scales,  are available
for the QCD fits. The extrapolation of these data to the 
LHC virtuality scales must involve, 
at present,  modeling of static nuclear effects, which are 
not controlled by the perturbative QCD.

\subsection{Monte Carlo event generator}
\label{subsec:MC}

The complexity of the interplay of the QCD and QED radiation 
and sophisticated  experimental procedures to derive
$W$-boson fluxes from the LHC data requires 
a dedicated Monte Carlo event generator.
The ultimate Monte Carlo generator should,  ideally,  include 
the full set of QCD and electroweak radiative corrections, 
re-summed in the phase-space regions where emission 
of soft radiation quanta is abundant, the phenomenological 
parameterizations of non-perturbative effects, such 
as the intrinsic transverse momentum of partons. It should be based  
upon  the most precise PDFs for nucleons and nuclei.  
Such a Monte Carlo generator is being developed at the 
moment, see e.g.~\cite{WINHAC:2003,Jadach:2003bu,EvolFMC:2004,EvolCMC1:2004,EvolCMC2:2004}.

In  the present study we use its present development version, 
the Monte Carlo event generator {\sf WINHAC}~\cite{WINHAC:2003,WINHAC:MC}.
At the  present  development stage,  
this generator incorporates only the leading-order process 
of creation of $W$-bosons,  
but includes already the EW radiative corrections in leptonic $W$ decays.
More precisely, collinear configurations of initial quarks are generated
from the PDFs,  while  the perturbative QCD effects are included through 
appropriate scaling violation. 
All  $W$-bosons produced by the present generator have  
$p_T^W=0$.
A set  of PDF parameterizations 
is provided through the {\sf PDFLIB} package~\cite{PDFLIB:2000}. 
 
Since,  at hadron colliders,  the $W$-bosons can be identified efficiently
only through their leptonic-decay channels, 
only leptonic  decays are currently implemented in {\sf WINHAC}.
The process of leptonic $W$ decays is described within the framework
of the Yennie--Frautschi--Suura exclusive exponentiation~\cite{yfs:1961},
where all the infrared QED effects are re-summed to the infinite order.
The residual non-infrared EW corrections are calculated perturbatively.
In the current version of the program the latter corrections are included up 
to ${\cal O}(\alpha)$.
{\sf WINHAC} went successfully through several numerical 
tests~\cite{WINHAC:2003},
and was also cross-checked with the independent Monte Carlo
program {\sf HORACE}~\cite{WINHAC-HORACE:2004}. 

The unique  merit of the {\sf WINHAC} event generator, for the 
studies discussed in this paper,  is that the $W$ production and decay
processes are described using the  spin amplitude formalism. 
The spin amplitudes for the $W$ production and the $W$ decay are calculated
separately. They correspond to all possible spin configurations of the
intermediate $W$-boson,  and of the initial and final-state fermions. 
The matrix element for the charged-current Drell--Yan process is
obtained by summing the production and decay amplitudes over the 
intermediate-$W$ spin states. The amplitudes are evaluated numerically 
for given particles four-momenta and polarizations. They can be
calculated in any Lorentz frame in which the corresponding particles
four-momenta are defined. For more detailed discussion see Ref.~\cite{WINHAC:2003}.
The advantage of using the spin amplitudes is that one can control the 
spin states, in particular the production of longitudinally and
transversely polarized $W$-bosons. Within this formalism one can also
include easily the effects of collisions  of the $W$-boson in hadronic matter
assuring the coherence of the mixed states of the $W$-bosons
throughout the collision and decay phases.

This Monte Carlo program allows to provide a first glimpse on the spectra 
and polarization of produced $W$-bosons, also in the presence of realistic 
experimental 
event selections. While the overall longitudinal spectra  of $W$s are well 
controlled, the current version of the program does not provide
information on their transverse spectra.
The latter can be described realistically by initial-state parton
shower algorithms, possibly matched with the NLO contributions 
to the hard process.
This will be worked out further 
in Refs.~\cite{EvolFMC:2004,EvolCMC1:2004,EvolCMC2:2004}.

The partonic distributions for the proton are taken from the {\sf PDFLIB} package 
-- in our studies we used the MRS~(G) parametrization ({\tt Ngroup=3} and 
{\tt Nset=41} in the {\sf PDFLIB} notation~\cite{PDFLIB:2000}). 
The partonic distributions for ion beams are also taken from {\sf PDFLIB}. 
They include the nuclear shadowing effects parametrized by the EKS 
group~\cite{EKS:1999}.
This parametrization is based upon the DGLAP extrapolation of the nuclear targets DIS data 
to the hardness scale of the $W$-boson production, and upon 
a phenomenological modeling of the static nuclear effects.

\subsection{Fluxes}
\label{sec:fluxes}

\subsubsection{Beam momentum spectra}
\label{subsec:beamspec}

\begin{figure}
\begin{center}
\setlength{\unitlength}{1mm}
\begin{picture}(160,150)

\put(0,78){\makebox(0,0)[lb]{
\epsfig{file=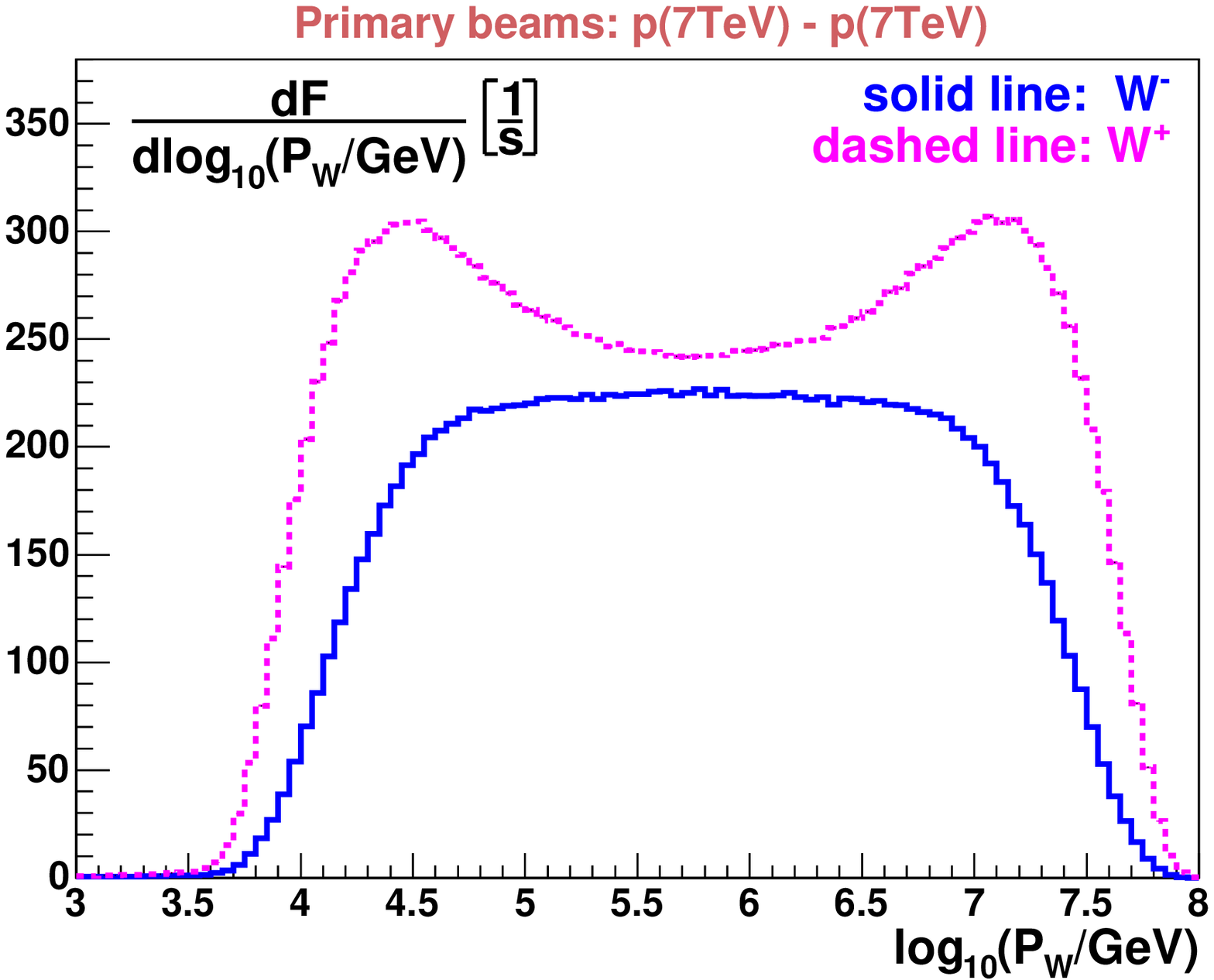, width=80mm,height=70mm}
}}

\put(0, 0){\makebox(0,0)[lb]{
\epsfig{file=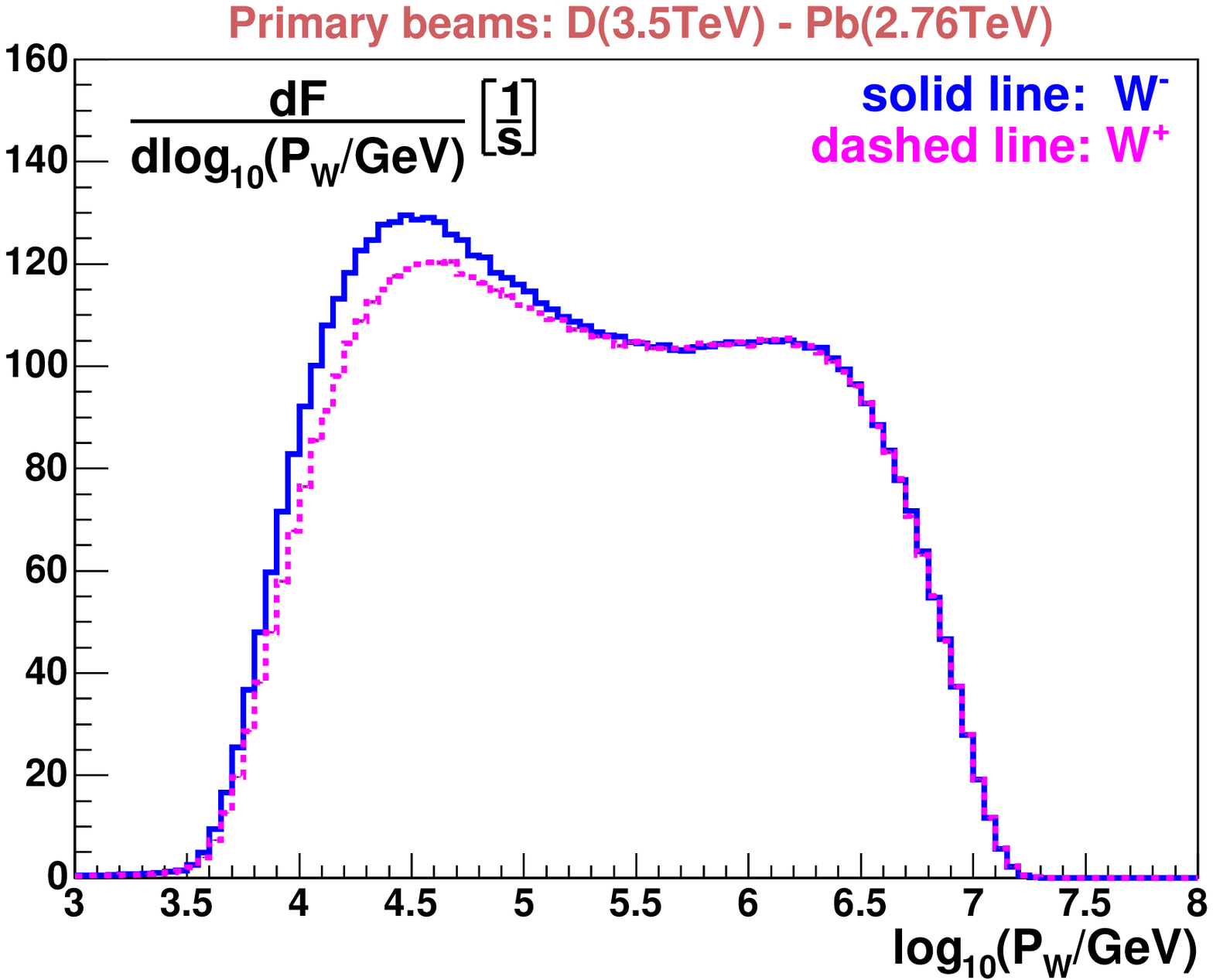, width=80mm,height=70mm}
}}

\put(75,78){\makebox(0,0)[lb]{
\epsfig{file=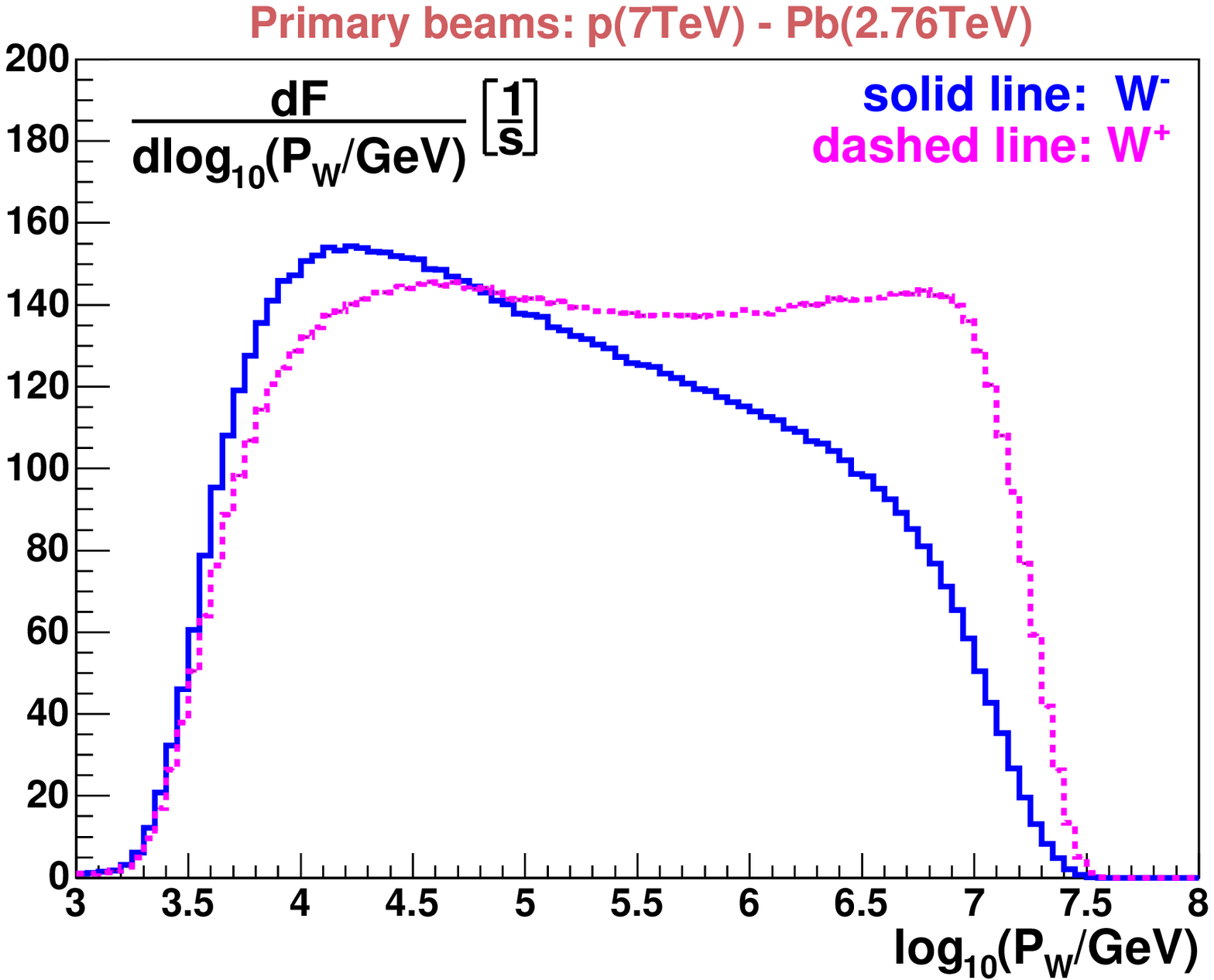, width=80mm,height=70mm}
}}

\put(75, 0){\makebox(0,0)[lb]{
\epsfig{file=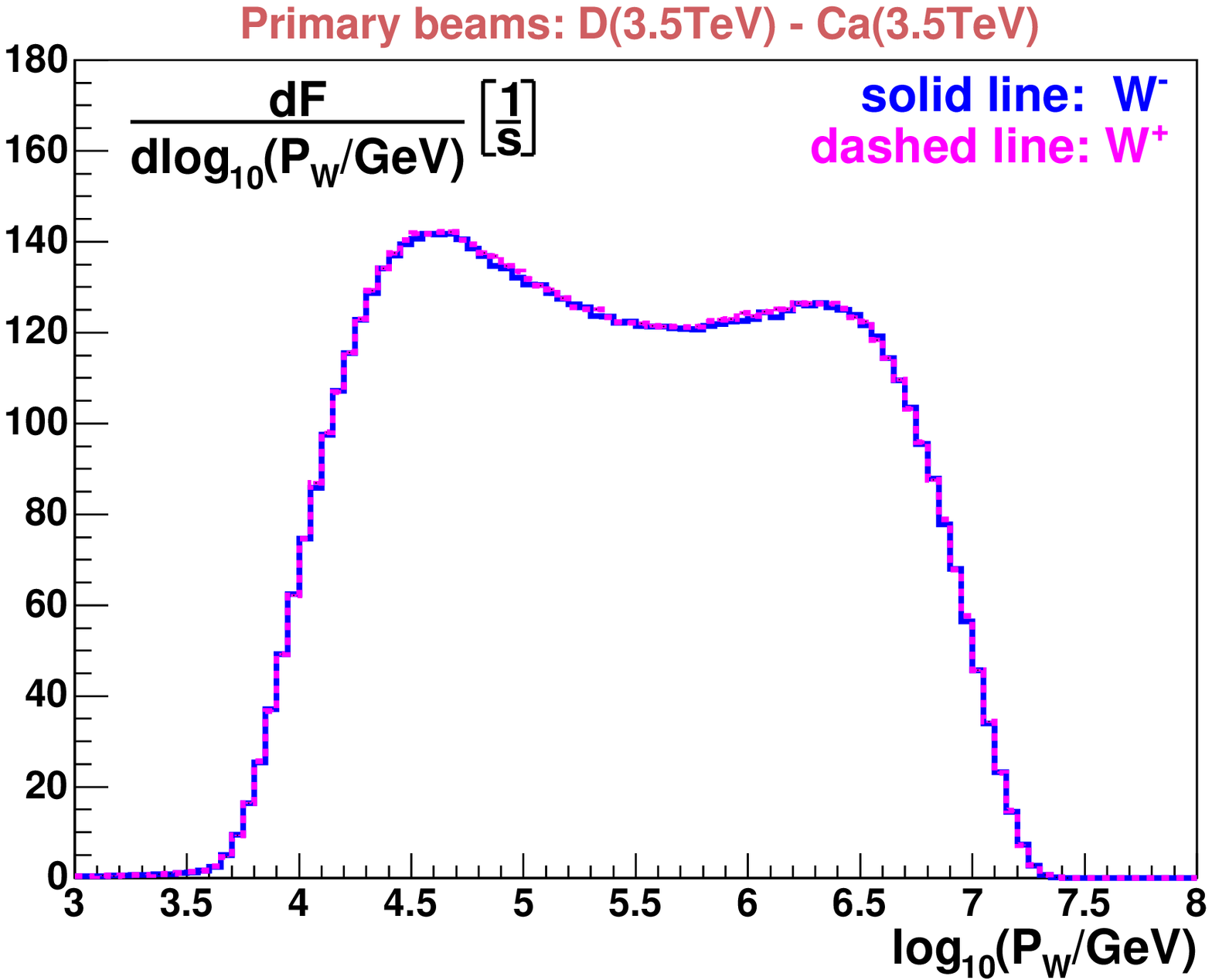, width=80mm,height=70mm}
}}
\put( 40,73){\makebox(0,0)[cb]{\bf (a)}}
\put(115,73){\makebox(0,0)[cb]{\bf (b)}}
\put( 40,-5){\makebox(0,0)[cb]{\bf (c)}}
\put(115,-5){\makebox(0,0)[cb]{\bf (d)}}

\end{picture}

\end{center}
\caption{\sf The fluxes  of the $W^{+}$ and $W^{-}$ beams for the 
         collisions of: 
        (a) $p$--$p$, (b) $p$--$Pb$, (c)  $D$--$Pb$ and (d) $D$--$Ca$, 
        for the luminosity of the primary collisions: 
        $L_{A_1 A_2} = 10^{34}/(A_1A_2)\, [{\rm cm}^{-2}\,{\rm s}^{-1}]$,
        as a function of the $W$-boson momentum in the 
        collinear $W$--nucleon  collision frame .}  
\label{pWspectra}
\end{figure}

The {\sf WINHAC} predictions for the fluxes of the $W$-bosons 
$  {\cal F} _{W}  ( p_n,p_A,p_W^{in} | A) $
produced in four LHC 
collision schemes: $p$--$p$, $p$--$Pb$, $D$--$Pb$,
and $D$--$Ca$  (where $D$ denotes deuteron) are shown 
in Fig.~\ref{pWspectra}. The fluxes are plotted 
as a function 
of the logarithm of  the $W$-boson momentum, 
$|p_W^{in}|$,  measured in the collinear $W$--nucleon  collision frame defined in 
Section~\ref{subsec:xsecs}.

The energies of primary beams  satisfy the equal magnetic
rigidity condition:
\begin{equation}
E_{\text{beam}} = 7 \;\frac{Z}{A}\; {\rm TeV},
\end{equation}
where $Z$ and $A$ are, respectively, the ion charge and its atomic number.
These plots are normalized  using the design luminosity for 
the $pp$ scattering: 
$L_{pp} = 10^{34}\,{\rm cm}^{-2}\,{\rm s}^{-1}$.  We have assumed 
the following  scaling 
of the luminosity  with the atomic numbers 
$A_1$ and  $A_2$ of the primary beams:
\begin{equation}
L_{A_1 A_2}= \frac{ L_{pp} }{ A_1 A_2}\,.
\label{scaling} 
\end{equation}
Such an assumption is realistic for collisions in which one of 
the colliding beam has a very small $A$.
In such a case  the beam--beam luminosity 
limits are less important than the limits driven by the bunch parameters,  and
by their collision frequency. In addition such a scaling is  practical for  
studying  the collisions of the point-like constituents of the beam particles. 
The luminosity scaled according to  eq.~(\ref{scaling}) assures the
$A$-independent luminosity of the parton--parton collisions in the 
limit of a dilute gas of partonic clouds.

The fluxes of the beam particles  vary in the range from  $200$ ($p$--$Pb$) 
to 370 ($p$--$p$)  $W$-bosons per second.  
It is interesting to note that these fluxes are similar to  
the fluxes of secondary hadrons 
used in the in the first bubble-chamber experiments at CERN.
The beam is a Wide-Momentum-Band  beam. Its momentum 
spectrum extends over four orders of magnitude,  from $10$~TeV 
to $100$~PeV, 

The  momentum distribution of the $W$-bosons produced in  proton--proton 
collisions, shown in Fig.~\ref{pWspectra}a, exhibits an asymmetry 
of  the spectra  of positively and negatively charged
$W$-bosons. This is a direct consequence of 
a net  excess of the $u$-quarks with respect to the $d$-quarks.
It is particularly large  in the small and in the large 
$\log_{10} {|p_W^{in}|}$ regions, 
where the $W$-bosons are produced mostly in the collisions 
of the valence quarks of the incoming beam particles.
The charge asymmetry reflects directly the ratio of the $d$-quark 
and the $u$-quark densities in this region. 
The momentum distribution of the $W$-boson flux
in the proton--lead collisions is  shown in Fig.~\ref{pWspectra}b.
The charge asymmetry changes sign for the ``small''-momentum $W$-bosons  
due to the excess of neutrons over protons
in heavy ions and the corresponding  excess of the $d$-quarks over the $u$-quarks. 
It disappears partially, as  illustrated in Fig.~\ref{pWspectra}c, 
for the deuteron--lead collisions
(in the large  $\log_{10} {|p_W^{in}|  }$ domain)  
and completely, as  illustrated in Fig.~\ref{pWspectra}d,
for the deuteron--calcium collisions. 

The size of the nuclear  effects in partonic distributions  (departure from a dilute-gas approximation)
is illustrated in Fig.~\ref{pWspectra}d, 
where  the momentum distribution 
of the $W$-boson flux is plotted for the primary collisions of iso-scalar beams 
having equal number of the $d$ and $u$-quarks.
The asymmetry of the low-momentum part and the 
high momentum part reflects the asymmetry in the nuclear corrections
of the two beam particles. 
This collision configuration is particularly  interesting  and 
will be discussed in detail in the forthcoming paper \cite{forthcoming}. 
In this configuration  the fluxes of the $W$-bosons  and those  
of the  $Z$-bosons can be directly related to each other.

The above plots illustrate 
the dynamic range in which  the momentum spectrum of the $W$-beam
can be tuned at the LHC. The knobs for such a tuning are:  
the choice of the atomic numbers of 
colliding particles,  and the selection of  
the charge of the $W$-boson. 
Such a tuning capacity, together with the capacity of tuning the energies
of primary beams  will play an important  role  in 
controlling the precision of unfolding the $W$-collision observables%
  \footnote{We would like to recall  that  varying the momentum byte and 
            the relative proportion of the pions and kaons
            generating the neutrino beams for the SPS neutrino collision program  was
            indispensable  for precise measurements of the neutrino--nucleon  and
            antineutrino--nucleon cross sections~\cite{Krasny-Bari, CDHS}.}.
The studies of internal consistency of
the momentum spectra  for  the positive and for negative $W$-bosons  
for variable atomic number  of the primary beams will be of important 
help in scrutinizing independently the PDFs  (both for the valence and the 
sea quarks) and the effects driving the energy dependence of the partonic 
emittance.

\subsubsection{Polarization}
\label{subsubsec:polar}

For the studies of collisions of the $W$-boson with the nucleon the natural 
choice of the spin quantization axis is the $z$-axis of  the 
{\it collinear $W$--nucleon collision frame} defined in Section~\ref{subsec:xsecs}.
In this reference frame the spin direction of the  
longitudinally  polarized $W$-boson 
is  perpendicular to the $W$--nucleon collision axis.

The polarization of the $W$-beam is a direct consequence of the 
$V-A$ coupling of the $W$-boson to quarks. If quarks were massless
and if they moved  collinearly with their parent hadrons then the $W$-bosons would
only be transversely polarized. The relative proportion of the  
 $\lambda =+1$ and the  $\lambda =-1$ fluxes would reflect 
the probability of the incoming quark to move along the $-z$ or $+z$
direction. 
Longitudinally   polarized  $W$-bosons ($\lambda = 0$)  are produced  
in the process of  annihilation of  massive quark(s)  and/or 
in the annihilation of  the quarks which 
are not collinear with their parent hadrons.
The relevant  experimental observable which allows to 
tune the  relative fluxes of the transversely and 
longitudinally polarized $W$-bosons
is the transverse momentum,  $p_T^{\text{recoil}}$, 
of the recoiled particles in  the $W$-production process.

\begin{figure}[!ht]
\leavevmode
\begin{center}
\setlength{\unitlength}{1mm}
\begin{picture}(160,70)
\put(0,0){\makebox(0,0)[lb]{
\epsfig{file=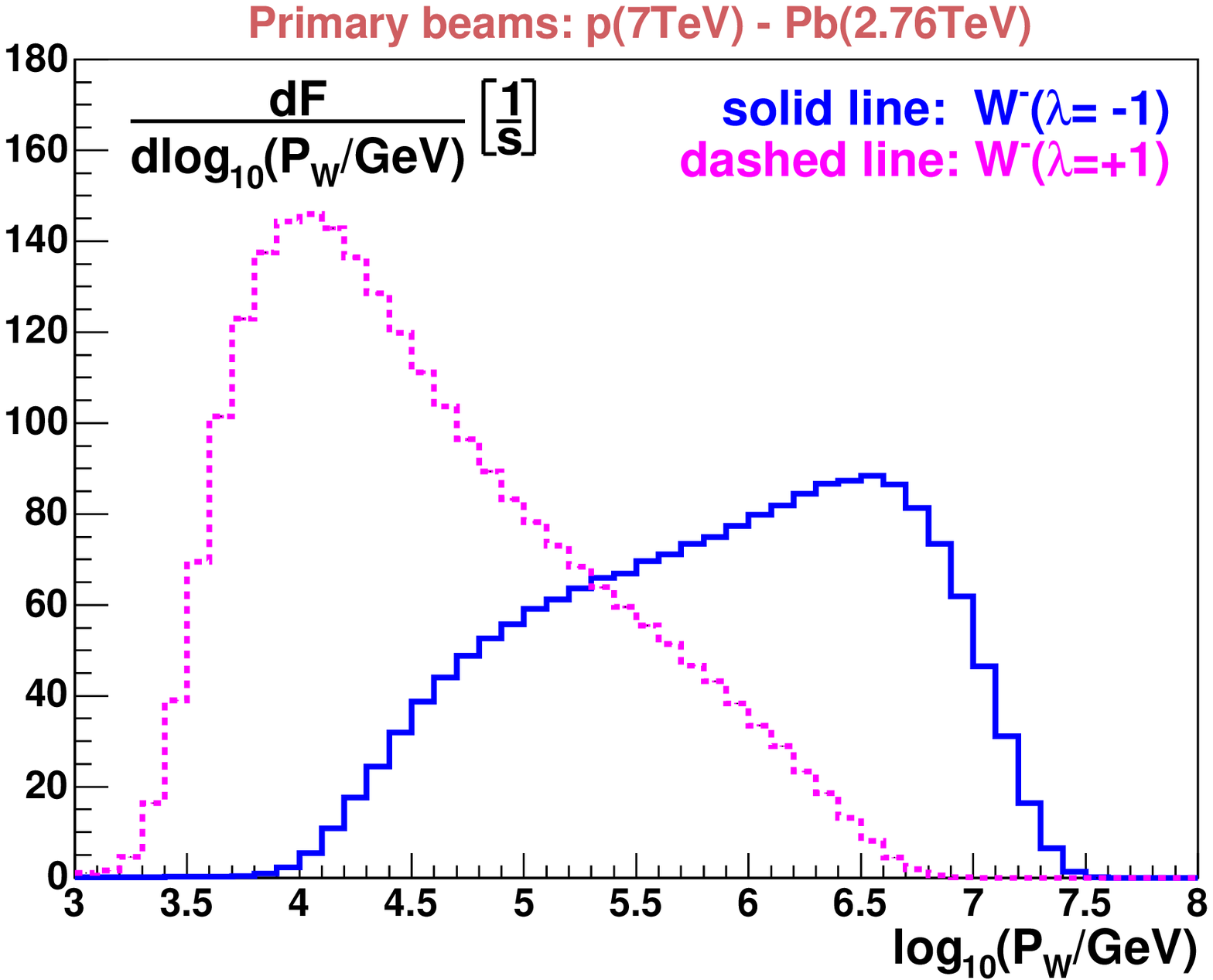, width=80mm,height=70mm}
}}
\put(75,0){\makebox(0,0)[lb]{
\epsfig{file=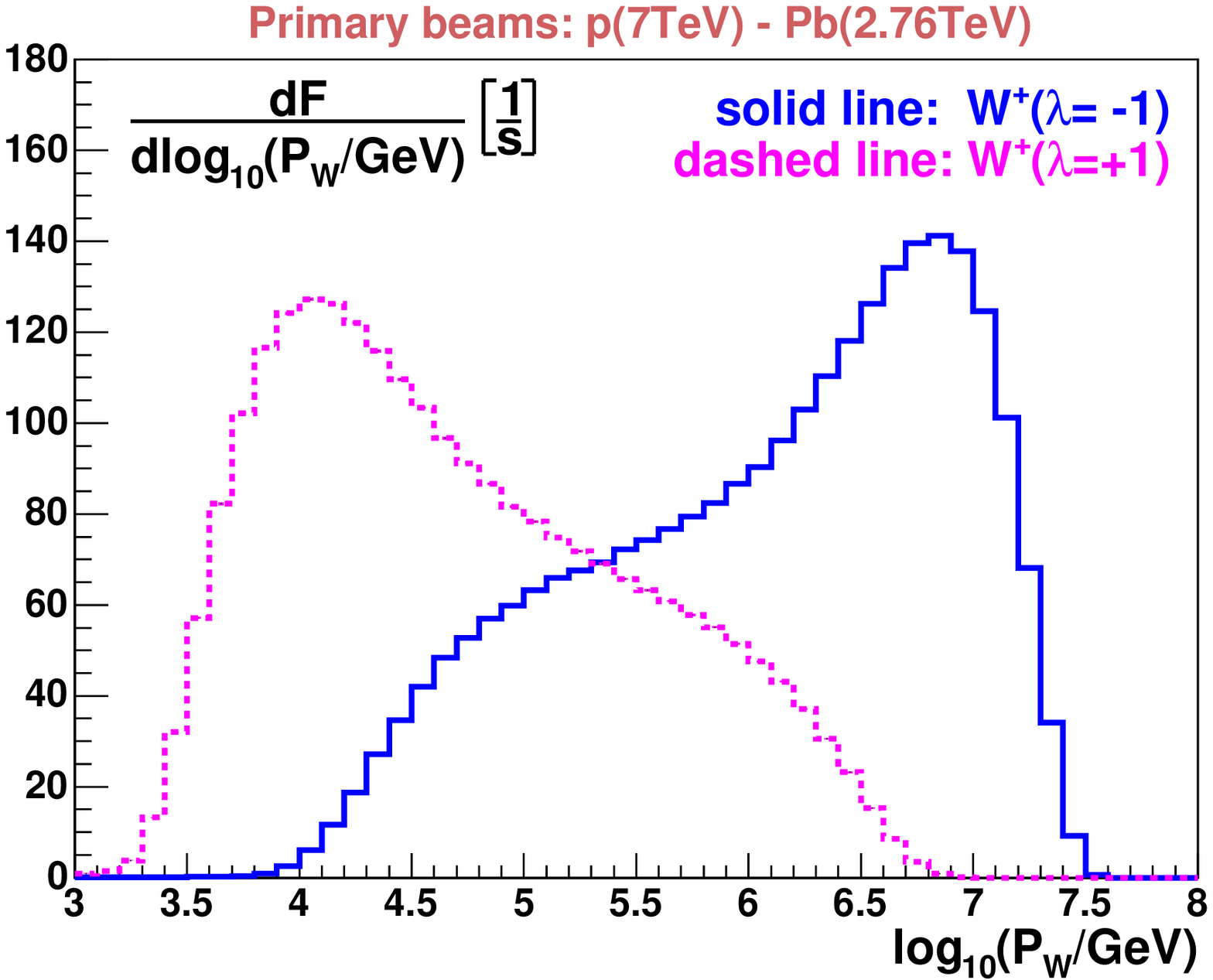, width=80mm,height=70mm}
}}
\put( 40,-5){\makebox(0,0)[cb]{\bf (a)}}
\put(115,-5){\makebox(0,0)[cb]{\bf (b)}}
\end{picture}
\end{center}
\caption{\sf The fluxes of the transversely polarized $W$-beams.}
\label{transverse}
\end{figure}
In Fig.~\ref{transverse} we show the  {\sf WINHAC} predictions for 
the polarization dependence of the 
$W$-boson beam fluxes:  
${\cal F}^{\lambda_{in}} _{W}  ( p_n,p_A,p_W^{in}  | A)$,  for the  primary collisions 
of the proton and  the lead-ion beams. The fluxes are  plotted as 
a  function of the beam momentum,  separately for the beams 
of the positively and the negatively charged $W$-bosons..
At the lowest  momentum  the $W$-beam has purely the $\lambda =+1$
polarization while at the highest one it has  the $\lambda =-1$ polarization. 
The relative proportion of the ${\cal F}^{1}_{W}$ and ${\cal F}^{-1}_{W}$ 
changes monotonically with the $W$-beam momentum. 
This plot illustrates the capacity of tunning the relative proportion 
of the  $\lambda =+1$ and  the $\lambda =-1$  fluxes within the full dynamical 
range by choosing the appropriate momenta of the primary beams and, 
to a certain degree, by choosing the charge of the beam  particles.

\begin{figure}[!ht]
\begin{center}
\epsfig{file=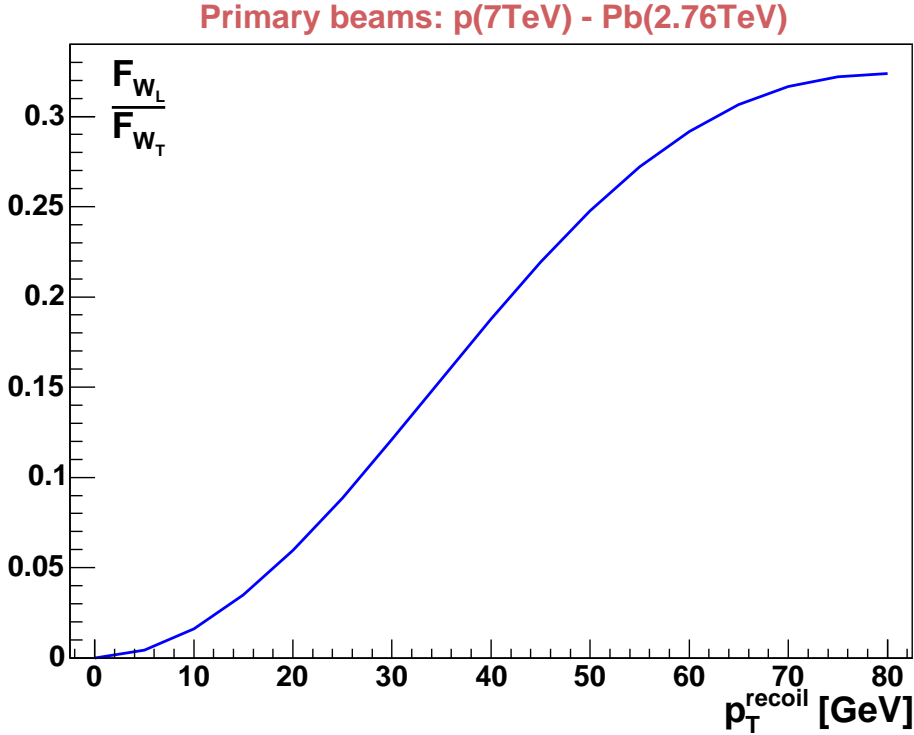,width=14cm}
\end{center}
\caption{\sf The relative flux of the longitudinally and transversely polarized
             $W$-bosons as a function of the $W$-boson recoil transverse momentum.}
\label{RW}
\end{figure}
In the  present  version of  {\sf WINHAC}  the 
NLO QCD processes  generating 
the transverse momenta of partons
are not included.
For the preliminary estimation  of the 
$p_T^{\text{recoil}}$ dependence of  the ratio of the   
fluxes of longitudinally and transversely polarized
$W$-bosons the effect of the transverse momentum of 
incoming partons was  emulated using the following kinematic model.
The transverse momentum of the $W$-boson was  generated
randomly and attributed with equal probability to the 
quark (anitiquark) of the nucleus and to the antiquark (quark)
of the nucleon. Since the quantization axis in the $W$-boson rest frame 
points to the direction of the nucleus, 
the $W$-boson may have, in this reference frame, the longitudinal polarization 
only if the transverse momentum 
is  attributed to the quark (antiquark) of the nucleus.
The ratio of the fluxes of the longitudinally and transversely polarized 
$W$-bosons was  determined using  the value of   
of the Wigner rotation angle -- the angle between 
the momentum vectors of the nucleus and of the quark
in the rest frame of the $W$-boson.   
The estimated ratio  is shown in Fig.~\ref{RW}.
This plot illustrates the correlation between the transverse momentum 
of the $W$-boson recoil particles  and the relative proportion of fluxes. 
Such a correlation allows to tune the relative
intensity of the longitudinally and transversely polarized $W$-beams.

\subsection{Experimental control of fluxes -- outlook}
\label{subsec:expflux}

The precision of controlling the fluxes of the $W$-bosons at the LHC  
will be gradually improved, before the start-up of the LHC operation,  by 
matching the  progress in detailed  understanding 
of the Tevatron $W$-production data with  the increasing 
precision of theoretical calculation of the $W$-production processes.
The ultimate precision which  will be achieved on the ``D-day'' of the
start-up of the LHC operation will, however, very likely be insufficient for
scrutinizing the collisions  of weakly interacting particles with hadronic  matter. 
Therefore,  one  needs to develop  dedicated measurement  methods  allowing 
to reduce the influence of  the $W$-boson flux uncertainty on the 
$W$-boson  collision observables.

Depending on  the required level of sensitivity to the $W$-beam
collision effects  the following three  methods of handling the $W$-boson 
fluxes are proposed in this paper:
\begin{itemize}
\item 
\it {the calculated flux} method,
\item
\it{the extrapolated flux} method, 
\item   
\it{the flux-independent} method.
\end{itemize}

The first method uses the 
calculated fluxes of the $W$-bosons.  
Initially, at the start-up of the LHC operation, 
its  sensitivity to the 
$W$-collision effects will be poor due to 
the $W$-flux shape and normalization uncertainties. 
The main effort will have to  be focused on 
diminishing the uncertainties  in  the non-perturbative inputs 
for the theoretical calculation of the  $W$-fluxes: in  the PDFs,  
and in the extrapolation of the transverse emittance of partons 
to  the LHC collision energy. 
The progress will most likely be driven  by  studies of  
the spectra  of leptons produced in collisions of the LHC  beams. 

A high-precision scrutinizing of  the PDFs could profit   
from the parasitic LHC electron beam, proposed  
recently in Ref.~\cite{ELECTRON}, to map the partonic distributions at the LHC.
If such a  beam is delivered to 
the LHC interaction points,  then 
the proton and the neutron PDFs could
be mapped  using the LHC detectors.  
Simultaneous measurements of the electron--proton(nucleus)
collision and the proton--proton(nucleus) collisions  
could  drastically diminish  the 
QCD-extrapolation and normalization uncertainties.
The  parasitic electron beam proposed in  Ref.~\cite{ELECTRON}
cannot, however be used to map the 
nuclear effects in the PDFs. These effects could be 
precisely measured at the HERA collider%
  \footnote{The nuclear program 
            for the HERA collider was developed and proposed in 1997 
            to the HERA community 
            as an alternative to the high-luminosity upgrade 
            \cite{Krasny-DESY,Krasny-GSI}.}. 
If not measured at HERA,
the PDFs of the nuclear bunches of quarks and gluons will thus relay 
on the scarce data and uncertain extrapolations -- until the 
eRHIC program will map them \cite{Krasny-eRHIC}.

In the second,   {\it extrapolated flux}  method  an  emphasis is put 
on understanding the variation  of the $W$-boson fluxes with the 
atomic number of the beam particles,  and on confining, as much as possible,
the dominant theoretical uncertainties of predicted $W$-boson fluxes within
the normalization factor --   the $A$-independent,   {\it dilute partonic system} flux.
Since the nuclear size  ($A$-dependent)  effects are present 
both in the $W$-boson fluxes and in the average path-length 
of the $W$-boson in hadronic matter they cannot in be 
resolved by measuring only 
the $A$-dependence  of the $W$-boson production rates. 
In the proposed   method   
the fluxes of 
the $W$-bosons are determined 
by measuring and fitting the $A$-dependence of the spectra  
of the charged lepton 
in the $\log{|p_l|} $ bins in the  {\it monitoring regions}.
The monitoring region  is defined 
by choosing: the energies of the primary beams,
the momentum of the $W$-boson,  and its polarization, 
such that  both the absorption and the inclusive cross sections 
for  $W$--nucleon collisions are  negligible (e.g.\ for transversely 
polarized $W$-bosons at low $W$--nucleon CMS energies).

The fluxes of the $W$-bosons and their  polarization can  be 
determined in the monitoring regions by measuring  
the angular distribution of the $W$-boson decay products.
The observed $A$-dependence of  the fluxes can   subsequently be used in 
unfolding the $W$-boson collision effects in a  configuration where 
they  become significant.
Such a  procedure allows to {\em factorize-out} and neglect 
a large fraction  of theoretical uncertainties of the $W$-fluxes.  It  
could be employed in experimental scrutinizing of the centre-of-mass energy 
dependence of the cross-section for collisions  of the longitudinally 
polarized $W$-bosons at the LHC-collider energies.

\begin{figure}
\begin{center}
\epsfig{file=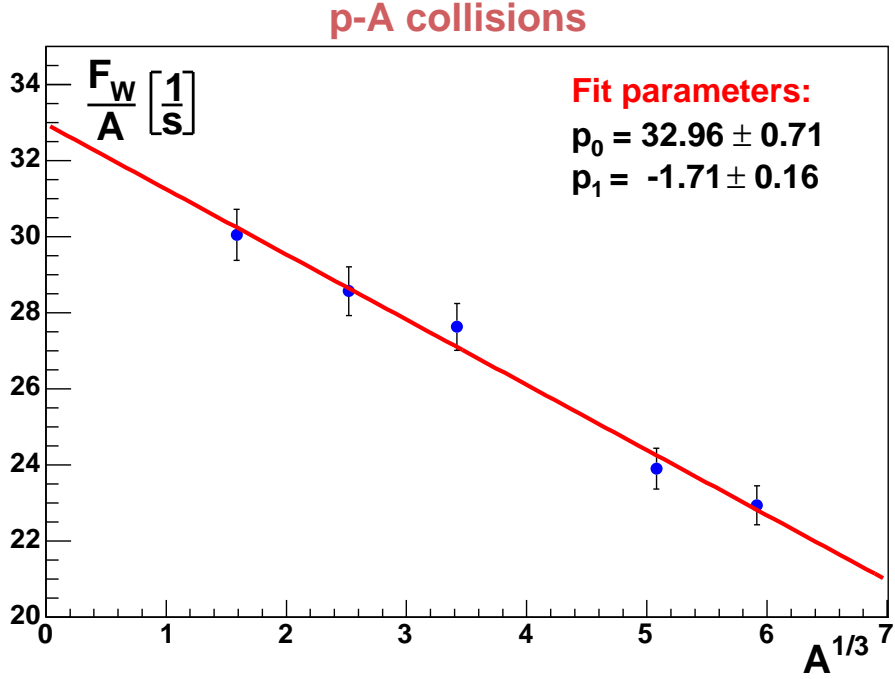,width=14cm}
\end{center}
\caption{\sf The fit of the A-dependence of the $W$-boson fluxes.}
\label{extrapol}
\end{figure}

The precision of such a procedure is illustrated in Fig.~\ref{extrapol}.
The $A$-dependence of the $W$-boson flux  is fitted 
in the monitoring region using the 
rates of the events containing  the $W$-boson decay signatures in the     
$p$--$He$, $p$--$O$, $p$--$Ca$, $p$--$Xe$ and $p$--$Pb$ collisions.
The  following form of the fit was chosen%
 \footnote{This is the simplest form of the fit which includes the 
            effects of shadowing for  the large momenta  of the $W$-bosons 
            ($\alpha \geq 0$),
             and the EMC effect for the small momenta of the $W$-bosons ($\alpha \leq 0$.)}:
\begin{equation}
 \frac{ {\cal F}_{W}}{A} =   {\cal F}_{0}\;  
                             \left(1- \alpha\, A^{\frac{1}{3}}\right).
\end{equation}
We have assumed that  
the point to point systematic normalization errors of the rates will be  kept below $2 \%$ 
and the $A$-dependent acceptance corrections will be  kept below  $1 \%$.
Such a precision can be  achieved already in the initial period of the 
LHC operation. 
The slope of the fit is determined,  with the precision of  $10\%$,
no matter what  theoretical precision 
of the overall normalization  of the {\it dilute partonic system flux}, 
$ {\cal F} _{0}  $,  can be achieved.

The key  point of the above  method is that 
only the relative point-to-point systematic errors are important in the fit of the 
$A$-dependence of the fluxes.
Since  the precision of the relative measurement
of the $W$-boson fluxes could  
be significantly improved in a dedicated machine configuration, 
e.g. by storing simultaneously 
or by frequent switching between various
iso-scalar nuclear beams,  this method has the  potential 
to be very precise.
In the  dedicated LHC running program 
with several ion beams 
a   $\sim 1 \%$  sensitivity to the $W$-collision effects
may be achieved in this method.

The  third, {\it flux-independent}  method minimizes the impact of the 
uncertainties in the $W$-boson flux for  specially designed  dedicated  measurements.
In this method, the effects  of propagation of the $W$-bosons in hadronic 
matter are related to the effects of propagation of other particles. 
For example, the collisions of the $W$-bosons in hadronic matter  
can be  monitored using  the monitoring sample of the $Z$-bosons.  
For the collisions of iso-scalar  nuclei (e.g. deuteron--calcium collisions),  for which the 
differences in the momentum distribution of the $d$-quark and the $u$-quark 
are negligible, the uncertainty 
of relative fluxes of the $W$ and $Z$-bosons can be reduced down  to the 
tiny effects due to their  mass difference. 
 In this collision configuration even a very small 
($\leq  0.1 \%$) anomalous effects in the 
propagation of $W$-bosons can be detected by 
using the $Z$-boson templates. This method will be 
discussed in more detail in the forthcoming paper~\cite{forthcoming}.

\section{Targets}
\label{sec:targets}

\subsection{Covariant and invariant quantities}
\label{subsec:covinv}

In the framework presented in Section~\ref{sec:femtoexp} the $W$-boson beam  targets  
are described by two functions: the average path of the 
$W$-boson in hadronic matter,  $\langle l_A(p_W^{in}, p_A)\rangle $, 
and the nuclear density, $\rho _A(p_A)$. In order to preserve the 
Lorentz-covariance of the formulae we have  kept explicitly their dependence 
upon the momentum of the nucleus,  allowing for the apparent  deformations
of its  volume and density in reference frames where the nucleus moves
with high velocity (e.g.\ in the $W$-boson rest frame).
The discussion of the control of the $W$-boson targets, presented in this
section, uses explicitly the fact that the product of these two functions 
is invariant with respect to Lorentz transformations. This is because 
the dependence of the nucleus density and of the effective path-length
of the $W$-boson upon the $\gamma$-factor of the Lorentz boost  cancel 
each other for spherically symmetric nuclei. 
This allows us to discuss the effective 
target length in the most convenient reference frame and use the formulas 
derived in this frame in arbitrary Lorenz reference frame,  provided that it will 
appear  in the expressions  together with the nuclear density.
The most convenient  reference frame  is the rest frame of the nucleus with the $z$-direction 
collinear with the direction of the incoming nucleon.

\subsection{$W$-boson path-length} 
\label{subsec:Wpathlen} 

The $W$-boson path-length  in hadronic matter
is determined by the relative position of the $W$-boson 
creation cell with respect to the centre of the nucleus and
by the $W$-boson momentum.
The  target length for the $W$-boson beam 
will thus vary on event-by-event basis.
The  event-by-event variation of  the target length 
cannot be controlled experimentally at the LHC%
  \footnote{At the LHC,   wounded nucleons of the nucleus are emitted 
            at small angles and cannot be detected.}.
What can be controlled experimentally  
is the $A$-dependent  average path-length of the $W$-boson in the  
nucleus of the atomic number $A$.
Since the nucleus-rest-frame $W$-boson momentum reaches  
the TeV--PeV range,  the observer traveling 
with the $W$-boson, views the nucleus as  a ``macroscopic''  classical 
object. The  quantum fluctuations of the its degrees of 
freedom are frozen  during the passage of the $W$-particle
allowing 
to describe the nucleus only in terms static nuclear parameters.

In Section~\ref{sec:femtoexp} we have implicitly  assumed, that the nucleon collides  
with the nucleus at the impact parameter $b=(b_x,b_y)= (0,0)$. 
The generalization of the formula (9) defining  the average path-length 
to the case where the nucleon arrives at the position of the nucleus with 
the uniformly distributed impact parameter is straightforward:  
\begin{equation}
\begin{aligned}
\langle l_A(  p^W_{in})  \rangle =  
\int db_x \int db_y \int dx_p \int dy_p \int dz_p\;
&
 l(x_p,y_p,z_p, p^W_{in},| A) \;  
\\
\times &
{\cal P}_{W} (x_p, y_p, z_p,b_x, b_y, p^W_{in}\, | A)\,,
\end{aligned}
\label{la}  
\end{equation}    
where $l(x_p,y_p,z_p, p^W_{in}\, | A)$ is the distance over which the 
$W$-boson produced with the momentum  $p^W_{in} $ travels  before leaving 
the nucleus of the atomic number $A$.  
The above formula is general but unpractical.
In the following we shall  simplify  it  and write it in a form which will 
allow us to express the average path of the $W$-bosons in terms 
of quantities which can be controlled experimentally.

Firstly, we recall  that
the quark (antiquark) of the nucleon projectile  
can be localized within the transverse distances of $L_{t} = 2 R_n$, 
where $R_n$ is the nucleon radius%
  \footnote{The contribution of diffractive, very small momentum-transfer $W$-production 
            ($\sqrt{t} \leq R_n/2$) can be safely  neglected here.}.
The antiquark (quark) of the target nucleus can be localized within  the 
transverse distances of $L_{t} =  2 R_A$,  where $R_A$ is the nucleus radius.
The transverse size of the $W$-boson formation cell (i.e. the 
region where, for given  $x_A$ 
and  $x_N$,  the $W$-formation process is  confined)  
is determined by  the overlap of the localization volumes  
of the quark (antiquark) of  the nucleon and antiquark (quark) the of  
the  nucleus.
The transverse position of the $W$-production cell is thus restricted to 
the region specified by the following boundaries: 
$  b_x - R_n \leq  x_p \leq b_x + R_n$, and 
$  b_y - R_n \leq  y_p \leq b_y + R_n$.
These conditions define the 
impact parameter and the $W$-boson momentum-dependent region in which 
the integration over $x_p, y_p$ will have to be performed. 
Since the transverse-size boundary is confined
to small distances (with respect to the nucleus size and the $W$-boson path-length),
 the integrals over $x_p, y_p$ can be 
calculated explicitly:
\begin{equation}
\langle l_A( p^W_{in})  \rangle =  
\int db_x \int db_y \int dz_p\;
 l(b_x, b_y, z_p, p^W_{in} | A) \; {\cal P}_W(b_x, b_y, z_p, p^W_{in} | A )\,.
\label{la1}  
\end{equation}    

Secondly, we note that the transverse momentum 
of the $W$-bosons is, for the LHC energies, significantly smaller 
than its nucleus-rest-frame longitudinal momentum. 
This allows us to ignore, in the calculation of the average path-length, 
the contribution coming from non-collinear  $W$--nucleus collisions. 
Technically,  this simplification  assures the cylindrical symmetry of the integral,   
allowing to factorize  the transverse and the longitudinal degrees of freedom.
In addition,  it allows us  
to drop the dependence upon the incoming $W$-boson  momentum $p^W_{in}$
in the path-length term   and 
upon the transverse momentum of the $W$-boson  in the 
probability term. 

This simplification results in the following expression:
\begin{equation}
\langle l_A(p^W_{in} )  \rangle =  
\int db_x \int db_y \;  \rho^n_A(b_x,b_y) 
\int  dz_p\; {\cal P}^b_W \left(z_p(p^W_{in}) \right) \,
l( b_x,b_y, z_p\, | A)\,, 
\label{la2}  
\end{equation}    
where $\rho^n_A(b_x,b_y)$ is the transverse density of nucleons 
in the nucleus, normalized to the integrated 
transverse density. The ${\cal P}^b_W $ is the probability  
of forming the $W$-boson
at the   longitudinal distance $z_p(p^W_{in})$  with respect to  the position 
of the centre of the nucleus for the fixed-impact-parameter 
nucleon--nucleus collision.
Its explicit  calculation requires the 
experimental knowledge of space-distribution 
(both in the impact parameter and in the longitudinal 
distance) of the quarks (antiquarks) involved in the production 
of the $W$-boson. The precise determination
of the  impact parameter dependence of the PDFs  was the  
principal  goal of  the nuclear program 
for HERA  ~\cite{Krasny-DESY,Krasny-GSI} and remains one of the  
target  of the future eRHIC experimental 
program at the BNL~\cite{Krasny-eRHIC}. Before such 
measurements are made,  the 
average path-length of the $W$-bosons 
can be controlled precisely  only in the 
kinematic domains  where the knowledge of the space-distribution
of partons  is not indispensable.

\subsection{Effective targets for $W$-beam}
\label{subsec:efftarg}

For the $W$-bosons produced at small momenta  in asymmetric collisions
of small-$x$ parton from the nucleon and large-$x$ parton from the nucleus,
the size of the $W$-boson formation cell: 
\begin{equation} 
  L_{l}(x_A) =  \frac{1}{x_A\, M_A}\; \leq 2R_n
\label{la3}
\end{equation}
is small and  the integrals in eq.~(\ref{la2})
can be calculated independently of the exact form  of the  ${\cal P}^b_W$.
This is because the $W$-boson formation cell is, in  such a case, localized 
both transversally and longitudinally within the volume of individual nucleons, 
and the total $W$-boson  path-length in nuclear matter is independent of 
the exact position of the $W$-boson creation point.
In this region  $\langle l_A(p^W_{in} )  \rangle$  is  independent of  the 
$W$-boson momentum  and can be expressed as:  
\begin{equation}
\langle l_A  \rangle =  
\int db_x \int db_y \; \rho^n_A (b_x, b_y)  \int dz_p \;
\frac{\sqrt{R_A^2 - b_x^2 - b_y^2}- z_p}{ 2\sqrt{R_A^2 - b_x^2 - b_y^2} }\,, 
\label{la2xx}  
\end{equation} 
where $R_A$ is the nucleus radius. 
This universal atomic-number-dependent quantity  representing 
the average target length for the $W$-bosons formed by the large-$x_A$ 
parton of the nucleus defines the {\it effective target length} for 
the $W$-boson beam. It can be tuned by choosing the atomic
number of the LHC ion beam%
\footnote{In the general case of an arbitrary Ioffe length the average 
          path-length is a function of the CMS energy of the $W$--nucleon 
          collisions $s_{Wn}$ and can be written as:  
          $ \langle l_A(s_{Wn})\rangle = K(s_{Wn})\, \langle l_A\rangle $,
          where $K(s_{Wn})$ varies between $1$, for the lowest energies, 
          and $2$, for the highest energies of the $W$--nucleon collisions 
          at the LHC.}.

\begin{figure}
\begin{center}
\setlength{\unitlength}{1mm}
\begin{picture}(160,70)
\put(0,0){\makebox(0,0)[lb]{
\epsfig{file=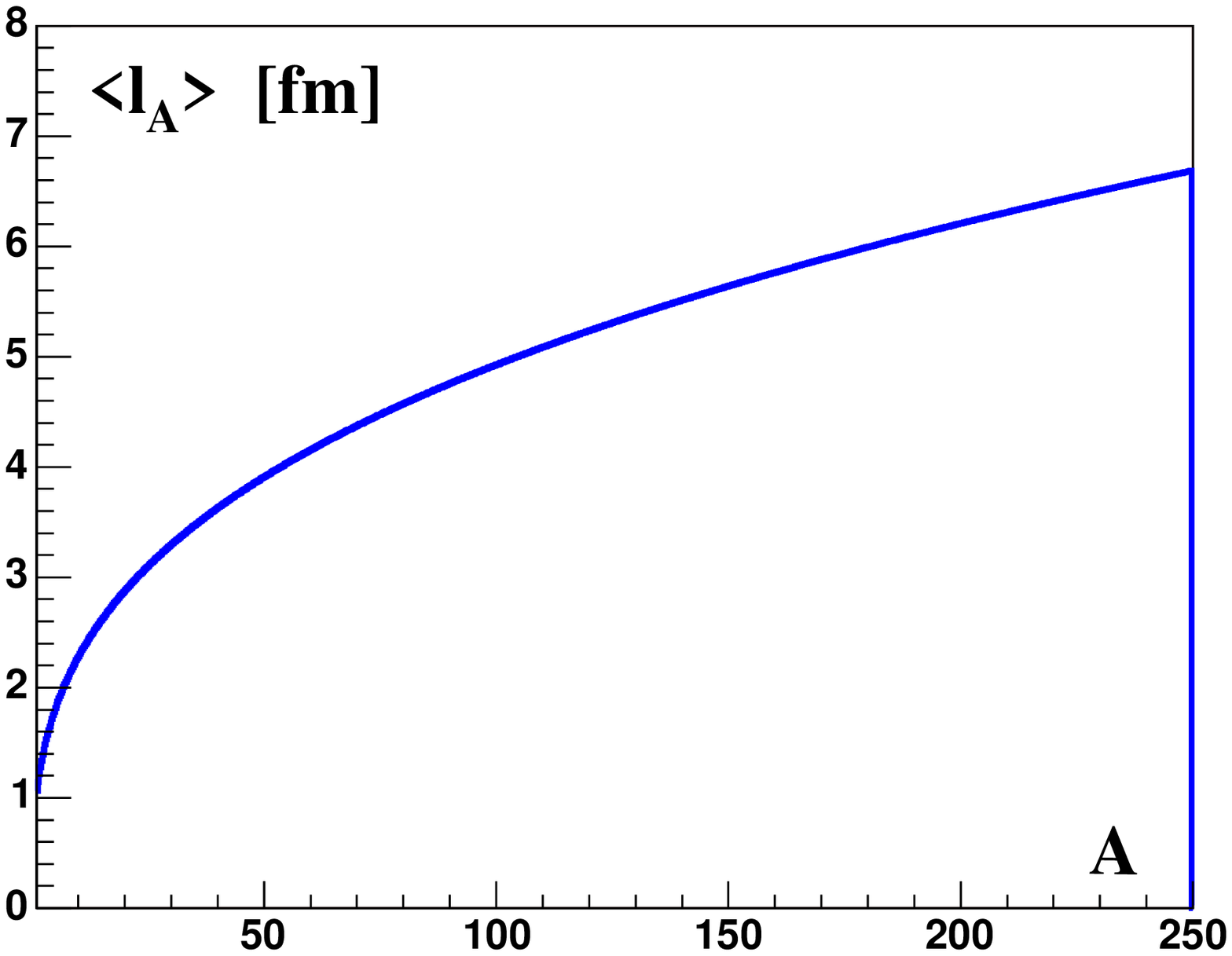, width=80mm,height=70mm}
}}
\put(75,0){\makebox(0,0)[lb]{
\epsfig{file=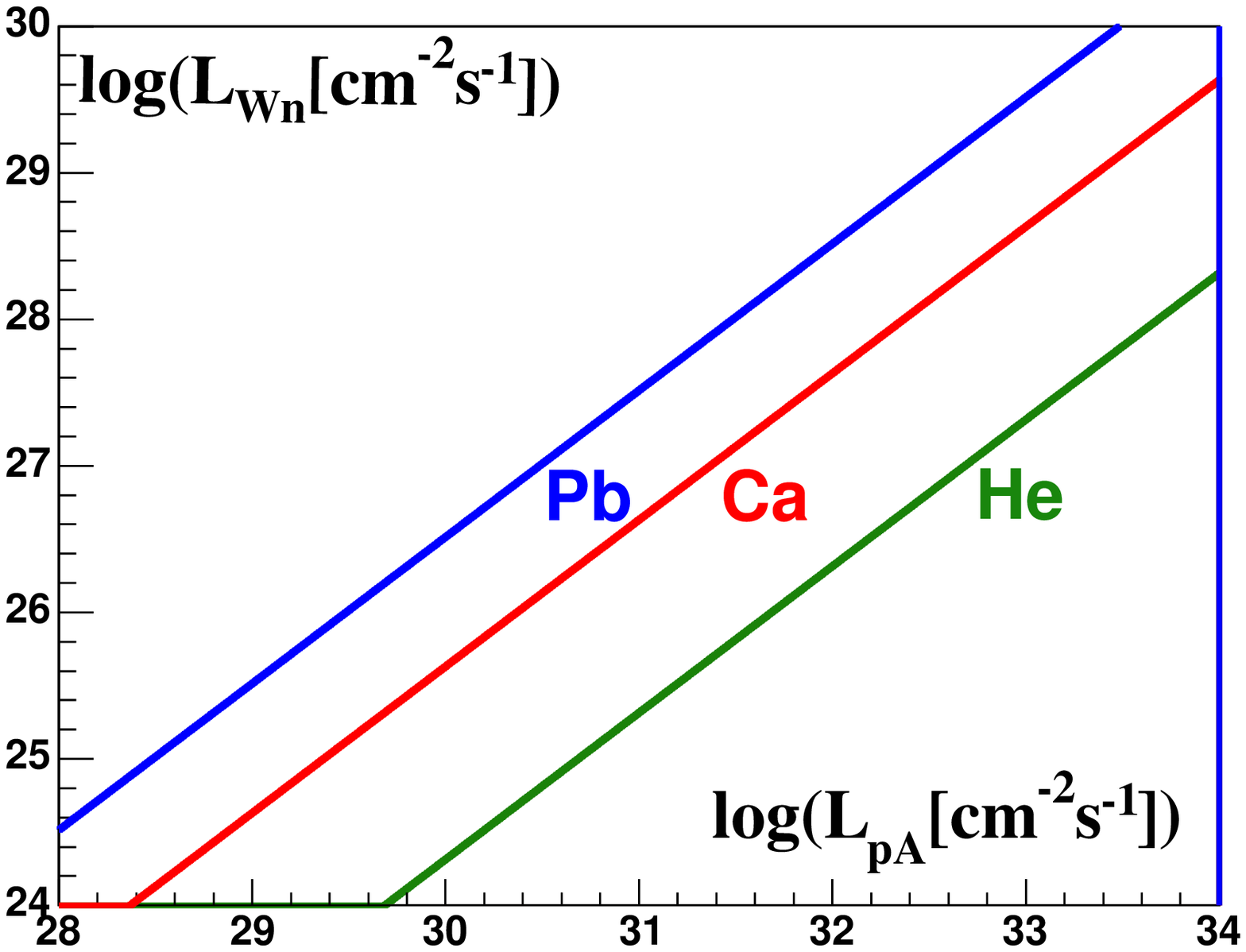, width=80mm,height=70mm}
}}
\put( 40,-5){\makebox(0,0)[cb]{\bf (a)}}
\put(115,-5){\makebox(0,0)[cb]{\bf (b)}}
\end{picture}
\end{center}
\caption{\sf The effective target length, (a), and the $W$--nucleon 
         luminosity, (b).}
\label{luminosity}
\end{figure}

In Fig.~\ref{luminosity}a we show the dependence of the average path-length
upon the  atomic number of the target nucleus  
for the $W$-bosons  satisfying the condition~(\ref{la3}).
The projected density was calculated using  the Saxon--Woods 
density~\cite{Saxon-Woods},  
and the nuclear radius was taken to be the Saxon--Woods parametrization radius.
The dynamic range of the average path-length  is confined to the
distances of  $ \sim 1$--$7$~fm, 
depending on the atomic number of the target 
nucleus.  

In order to demonstrate  the full equivalence of the  targets used in macroscopic 
scattering experiments with the nuclear targets for the femto-experiments,  
it remains to be demonstrated that the experimental control of the average 
length of the the $W$-boson beam target is equivalent to the event-by-event
control of the target-length in macroscopic experiments. 
This need not to be the case in general.
For the $W$-boson beam this requirement is likely to be fulfilled since
the $W$--nucleon collision cross section is expected to be  small enough that  
each produced $W$-boson could collide  at most once  within the nucleus:
$\sigma_{Wn} \ll 1$~fm$^2$. If this condition is fulfilled, then the sample of 
$N$ events in which the $W$-boson traveled the total 
distance $l_A(N) = \sum_{i=1,N}  l_A(i)$  
is,  in the limit of large $N$,  equivalent to the sample of $N$ events 
in which each produced  $W$-boson traveled the distance     $\langle l_A  \rangle$  given  
by eq.~(\ref{la2xx}).

\section{Luminosity}
\label{lumi}

The luminosity of the LHC $W$--nucleon collider, employing  the 
polarized $W$-boson
beam produced in the interaction of the primary beams of protons and 
ions of atomic number $A$,  can be expressed as:
\begin{equation}
 L_{Wn}^{\lambda_{in}}(s_{Wn})\, [{\rm fm}^{-2}\,{\rm s}^{-1}] = 
{\cal F}^{\lambda_{in}}_W  ( s_{Wn})\, [{\rm s}^{-1}]  
\times       \langle l_A(s_{Wn} )   \rangle  \,   [{\rm fm}]  \times \rho^n _A\, [{\rm fm}^{-3}]\,.
\label{lum} 
\end{equation}
In this formula we have written explicitly the units and expressed 
the kinematic observables in terms of the Lorentz-invariant centre-of-mass 
energy of the $W$--nucleon collisions.   
The flux of the $W$-boson in the polarization state $\lambda_{in}$  
enters the equation in the surface-integrated form. 
This simplification was  possible owing to the large $\gamma$-factor 
of the Lorentz-boost from the $W$-beam particle rest frame 
to the nuclear rest frame%
   \footnote{The lower boundary of the Lorentz $\gamma$-factor 
             depends upon the angular acceptance 
             for the $W$-decay products of the LHC detector. 
             In the case of  the ATLAS (CMS) detector 
              $\gamma \geq 100$.}, 
which reduces  the apparent  divergence  of the $W$-beam 
in the target rest frame.
The centre-of-mass energy of the $W$--nucleon collisions 
is unambiguously determined by the longitudinal momentum of the $W$-boson 
in the {\it collinear $W$--nucleon collision}  frame because the Fermi momenta 
of the nucleons within the nuclei can be neglected at the LHC energies.
Similar formula could also be written for the luminosity 
of $W$--parton collisions by 
replacing the distribution of the nucleons within the nucleus  $\rho^n _A$
with the distribution of the partons within the nucleus $\rho^p_A$ and by 
recalculating the formula~(\ref{la}) for the partonic density  
rather than nucleon  density. 
This  would preclude that $W$-bosons interact with hadronic  matter  only by 
a direct coupling  to its  point like constituents. 
Such an assumption, even if quite natural, will have to be 
tested  experimentally at the LHC. 

In order to asses the statistical sensitivity  of the $W$--nucleon collider 
for exploring the interactions of $W$-bosons with hadronic matter, 
we rewrite the formula~(\ref{lum}) by expressing the $W$-boson flux 
in terms of the luminosity of the proton--nucleus
collisions $L_{pA}$: 
\begin{equation}
 L_{Wn}^{\lambda_{in}}(s_{Wn})  =  L_{pA} \; 
 {\cal \sigma} ^{\lambda_{in}}_{pA \rightarrow  W+X}  (s_{Wn}) \;  \langle l_A(s_{Wn} )   \rangle  
 \; \rho^n _A.
\label{lum1} 
\end{equation}
The dependence of the total $W$--nucleon collision  luminosity, 
summed over all $W$-boson polarization states,  
upon the nucleon--nucleus collision  luminosity is shown 
in Fig.~\ref{luminosity}b for  three nuclear beams: 
helium, calcium and lead, and for the $Wn$-collision energies 
satisfying the condition of eq.~(\ref{la3}). 
This plot illustrates what values of luminosities  of 
the proton--helium, proton--calcium and proton--lead 
collisions need to be achieved  for the  required $W$--nucleon collision  
luminosity.  If the luminosity of  the 
$p$--$A$ collisions decreases with the  atomic number of the nucleus as   
$L_{pA} \sim  L_{pp}/A$, then 
the $p$--$Pb$ collisions are the most effective in  reaching  the highest 
possible sensitivity
to the $W$--nucleon collision effects. If it will decrease  faster,
e.g.\ like   $L_{pA} \sim L_{pp}/(A\,Z)$, then
the beams of lighter nuclei may turn out to be more effective.
Given the expected performance of the LHC collider for the  
proton--nucleus collisions,  the sensitivity 
to the $W$--nucleon cross sections down to the level of $100$~nb 
can be reached.
The systematic  sensitivity will be  determined entirely by the 
quality of the dedicated analysis methods capable of  reducing the overwhelming background to the 
$W$-boson collision effects due to conventional partonic processes.

\begin{figure}[!ht]
\begin{center}
\setlength{\unitlength}{1mm}
\begin{picture}(160,70)
\put(0,0){\makebox(0,0)[lb]{
\epsfig{file=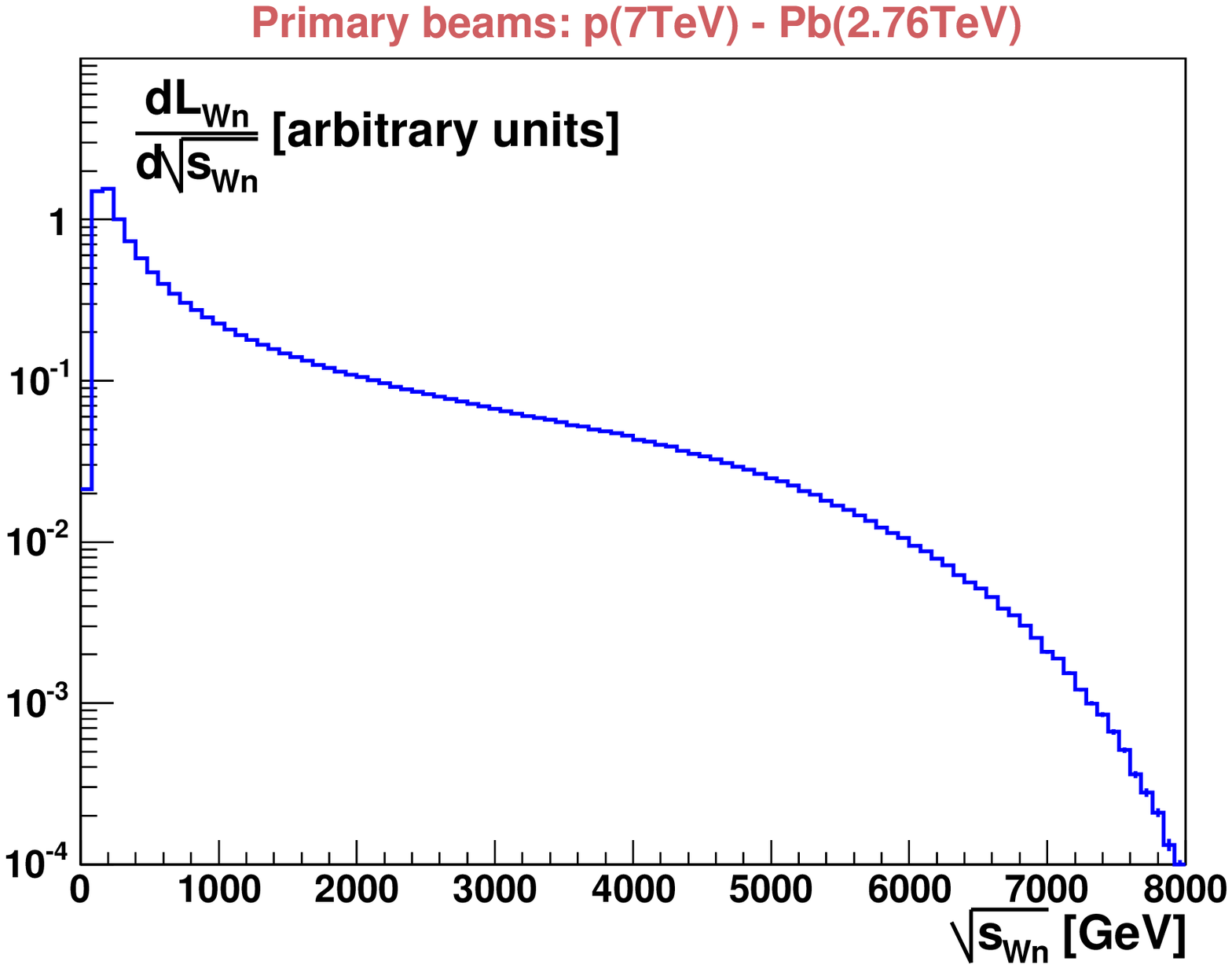, width=80mm,height=70mm}
}}
\put(75,0){\makebox(0,0)[lb]{
\epsfig{file=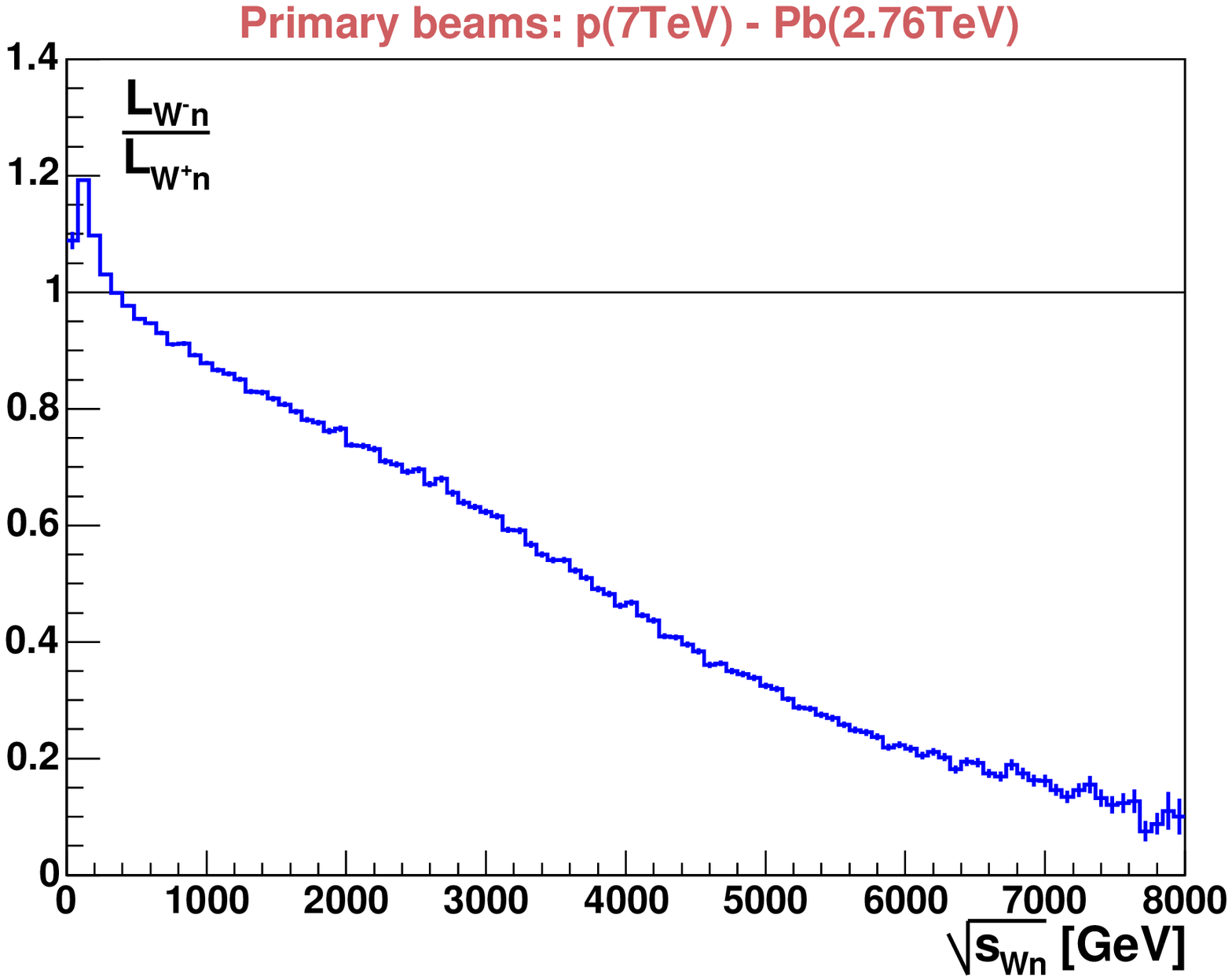, width=80mm,height=70mm}
}}
\put( 40,-5){\makebox(0,0)[cb]{\bf (a)}}
\put(115,-5){\makebox(0,0)[cb]{\bf (b)}}
\end{picture}
\end{center}
\caption{\sf The distribution of the $W$--nucleon CMS energy, (a), 
         and ratio of the collision-energy spectra for the $W^{+}$ 
         and $W^{-}$ beams, (b).}
\label{spectra}
\end{figure}

One of the merits of the $W$-beam at the LHC collider is that it is a 
broad-momentum-band beam. It will give access to a broad range 
of the $W$--nucleon collision energies. The spectrum of energies of the 
$W$--nucleon collisions is shown in Fig.~\ref{spectra}a. 
The accessible energies extend up to the TeV range -- the region 
where collisions of the $W$-bosons, in particular the longitudinally 
polarized $W$-bosons, are expected to shed light on the mechanism 
of the electroweak symmetry breaking.

In Fig.~\ref{spectra}b  we show the ratio of the energy spectra 
for positively and negatively charged $W$-bosons.
This plot illustrates the merit of having simultaneous beams of positively 
and negatively charged $W$-bosons for exploring  the energy dependence 
of the $W$--nucleon inclusive cross section.  
Even if the centre-of-mass energy of the $W$--nucleon 
collision is not directly reconstructed, the $W^+$ and $W^{-}$ 
spectra   are sufficiently distinct to look for 
the anomalies in energy dependence of the cross section, 
in the abnormal effective-target-length  dependence of the charge 
asymmetry of the final-state $W$-bosons.

\section{Outlook}
\label{outlook}

The luminosity  the $W$--nucleon  
collisions that  can be reached  at the LHC
may  not be high enough to explore the rare $W$-boson collisions (e.g.\ 
the $W$--$W$ collisions). The sensitivity to such processes 
is inferior with respect to the canonical proton--proton (nucleon--nucleus)   collisions, 
where the process of the virtual $W$-boson creation and its subsequent hard 
collision takes place 
in a single space-time volume%
\footnote{The rare, hard $W$--nucleon collisions can be considered
in this context as higher-twist processes in nucleon--nucleus collisions.}. 
The $W$--nucleon collisions  will be optimal, for  exploring
``soft'' $W$-boson collisions with hadronic matter, 
in particularly for  investigating  the  processes sensitive to the polarization of the 
initial $W$-boson.
Generic  studies of such collisions may lead to surprises  which are not  
present in the inventory lists of the canonical search scenarios.

The $W$--nucleon collision program at the LHC could start 
from the analysis of the electroweak-boson  production in 
the proton--lead collisions. This  configuration  is  already foreseen 
in the LHC collider program. Such  studies could be 
considered as a ``femto-detector R\&D'', having as a  goal to   
demonstrate the  feasibility of extraction of the $W$--nucleon collision 
signals in the proton--lead collisions.
If such a demonstration is made and the need for a generic 
exploration of  the $W$-boson collisions at the LHC energies 
arise (either because of confirmation or because of rejection of the canonical
scenarios of the electroweak symmetry breaking),
the configuration phase of the femto-experiment could follow. 
Configuring  the femto-experiment would consist
of choosing the optimal nuclear species, beam energies 
and the running time in each of the LHC collider setting 
for the highest sensitivity to the $W$--nucleon collision effects.

\section{Concluding Remarks}
\label{conclusions}

In this paper we have presented a generic, experimental  method of exploring the processes
of collisions of the $W$-bosons with hadronic matter in  the TeV-energy 
range. We have shown that the collisions
of the standard LHC beams can be treated as an incoherent source of the
polarized, wide-momentum-band  $W$-boson beam.
Such a beam can be used to study the 
collisions of the $W$-bosons with nucleons in analogous way 
to the use of the muon beam in the fixed-target experiments. 
We  have proposed  a coherent  data analysis framework allowing 
to relate both the spin-independent and the spin-dependent $W$--nucleon collision observables to the  
event rates measured in collisions of the primary LHC beams.
We have discussed the unfolding method  of the $W$--nucleon collision cross sections
and proposed the experimental methods of controlling the spectra and  the fluxes 
of the $W$-boson beam, 
its polarization,  and the effective lengths of the $W$-boson targets.  
The $W$--nucleon luminosities that can be achieved at the LHC reach the values
of $L= 10^{28}\, {\rm cm}^{-2}\,{\rm s}^{-1}$. 

In this paper we have not discussed the selection methods for filtering  out the 
signals of the $W$-boson collisions from the background 
of ordinary nucleon-nucleus collisions. This subject will be a addressed in \cite{forthcoming}.

The main purpose of this paper is  to draw attention 
to a very important and unique
merit  of the LHC nuclear beams  
for the  experimental  program aiming to explore, 
in the model independent way,   
the collisions of electroweak bosons
at the TeV energy scale.
Such a program can be uniquely  realized at the LHC collider owing to 
the versatility of its running modes and beam particle species.
It can be realized neither at the lepton-beam colliders nor at the 
Tevatron collider.
The luminous nuclear beams provide the  indispensable research tools for such a program.  
While their merits  for
the QCD sector of the Standard Model, in particular for  searches of the quark-gluon plasma,  
are widely accepted,  their merits for studies of the electroweak sector
of the Standard Model have, up to our best knowledge,  not been 
put so far into sufficiently bright light. This paper could be considered as an
attempt to fill in  this gap by showing  the unique role of  
the LHC collision scheme in which the light ions (protons)
are colliding with the heavy ions.

\vspace{5mm}
\noindent
{\large\bf Acknowledgements}
\vspace{3mm}

\noindent
We would like to thank  Y. Dokshitzer, F. Dydak and A. Bia{\l}as for useful discussions.



\end{document}